\documentclass[twocolumn]{aastex6}

\usepackage{color} 

 \newcommand{\co}{${\rm^{12}CO}\,$} 	
 \newcommand{\tco}{${\rm^{13}CO}\,$} 	
 \newcommand{\eco}{${\rm C^{18}O}\,$} 	
\newcommand{\kms}{km s$^{-1}$} 
 \newcommand{\1}{($J=1-0$)} 

\shorttitle{IR cluster}
\shortauthors{T.S. et al.}

\begin{document}


\title{A Statistical Study of Massive Cluster-Forming Clumps}


\author{
Tomomi Shimoikura\altaffilmark{1}, Kazuhito Dobashi\altaffilmark{1}, 
Fumitaka Nakamura\altaffilmark{2,3}, 
Tomoaki Matsumoto\altaffilmark{4},
and Tomoya Hirota\altaffilmark{2,3}
}

\affil{\scriptsize{\rm $^1$ ikura@u-gakugei.ac.jp}}
\altaffiltext{1}{Department of Astronomy and Earth Sciences, Tokyo Gakugei University, Koganei, Tokyo  184-8501, Japan}
\altaffiltext{2}{National Astronomical Observatory of Japan, Mitaka, Tokyo 181-8588, Japan}
\altaffiltext{3}{Department of Astronomical Science, School of Physical Science, SOKENDAI (The Graduate University for Advanced Studies), Osawa, Mitaka, Tokyo 181-8588, Japan}
\altaffiltext{4}{Faculty of Sustainability Studies, Hosei University, Fujimi, Chiyoda-ku, Tokyo 102-8160, Japan}


\begin{abstract}
We report results of the observations of 15 regions in several molecular lines for a statistical study of massive cluster-forming clumps. We identified 24 clumps based on the \eco ($J=1-0$) data obtained by the NRO 45 m telescope, and found that 16 of them are associated with young clusters. The clumps associated with clusters have a typical mass, radius, and molecular density of $\sim1 \times 10^3$ $M_\sun$, $\sim0.5$ pc, $\sim1 \times 10^{5}$ cm$^{-3}$, respectively. We categorized the clumps and clusters into four types according to the spatial coincidence of gas and star density, and discussed their evolutions: Clumps without clusters (Type 1), clumps showing good correlations with clusters (Type 2), clumps showing poor correlations with clusters (Type 3), and clusters with no associated clumps (Type 4). Analyses of the velocity structures and the chemical compositions imply that the clump + cluster systems should evolve from Type 1 to Type 4. We found that some of the Type 2 clumps are infalling on the clump-scale to form clusters at the clump center, which should commonly occur in the beginning of cluster formation. Interestingly, all of the identified Type 1 clumps are likely to be older than the Type 2 clumps in terms of chemical compositions, suggesting that they have been gravitationally stable for a long time possibly being supported by the strong magnetic field of $\gtrsim 1$ mG.
Type 1 clumps younger than the observed Type 2 clumps should be very rare to find because of their short lifetime.

\end{abstract}


\section{INTRODUCTION} \label{sec:intro}

Young clusters with ages of a few Myr are deeply embedded in their natal molecular clumps. 
These clusters can be easily identified in near-infrared images, 
and we call such young clusters IR clusters.
Although it is well known that stars are often born in clusters \cite[e.g.,][]{ladalada,Gutermuth2009}, 
 formation and evolution processes of clusters still remain unclear. 
Observational studies of IR clusters and their natal clumps provide 
a key to understand the formation and evolution of clusters and their environments.

Cluster-forming clumps tend to exhibit complex velocity structures with multiple velocity components.
Several hypotheses have been proposed to explain the velocity fields of cluster-forming clumps:
For example, it is possible that the clumps can be affected by an expanding H{$\,${\sc ii}} region \cite[e.g.,][]{Shimoikura2015}
or by collisions of smaller clumps and/or filaments \cite[e.g.,][]{Higuchi2009,Dobashi2014}.
\cite{Peretto,Peretto2007} found two velocity components in the massive cluster-forming clump NGC2264C,
for which they suggested a large-scale collapsing motion of the clump.
Similarly, we investigated the velocity structure of the cluster-forming clump S235AB,
and found that the entire clump is infalling with rotation toward the clump center
where the most massive star in the cluster is forming \citep{Shimoikura2016}.

How the velocity structures of the natal clumps evolve along with the cluster formation?
Considering the results we obtained for the clump S235AB \citep{Shimoikura2016}, we would expect 
that the clumps forming IR clusters should experience the clump-scale infalling motion with rotation 
in an early stage of cluster formation.
In order to confirm if such a dynamical infall is a common phenomenon for the cluster-forming clumps,
we need to carry out a statistical study with a rich sample of massive clumps with young clusters.

We also note the fact that there is a large variation in morphologies of the IR clusters and their natal clumps,
which is believed to represent various evolutionary stages of cluster formation \cite[e.g.,][]{ladalada}.
However, the quantitative relationships between the properties of
clusters (e.g., age, size, number of constituent stars)
and their natal clumps (e.g., mass, density, and velocity dispersion)
still remain unclear. The relationships should change along with the evolution time,
but it is generally difficult to assess the absolute ages of the natal clumps.
There are, however, some molecules (e.g., CCS) for which the time variations
of the abundances have been calculated \citep[e.g.][]{Suzuki1992},
and they can be used as a measure of the clump ages.
In order to study the relationships between the clusters and natal clumps,
it is therefore very important to investigate systematically not only
the physical properties but also the chemical compositions
of the clumps.

In this study,
we conducted systematic observations of 15 dust condensations
selected from a catalog of dust condensations compiled by \cite{Dobashi2011}.
We used the NRO 45 m telescope to observe the dust condensations with different tracers.

In the C$^{18}$O data from the 45 m telescope, the condensations can be divided 
into smaller clumps. 
In one of the clumps named S235AB corresponding to No.4423 in
the 15 dust condensations, we detected the global infalling motion
with rotation as mentioned in the above.
We show the results for S235AB in detailed in a separate paper \citep{Shimoikura2016}.
In this paper, we present the results for all of the clumps we identified.

Our primary goals in this paper are 
(i) to find differences between the clumps with and without IR clusters and 
to investigate how and under what physical conditions the cluster formation can occur
mainly by using the \eco data as a tracer of kinematics,
and
(ii) to infer the evolutionary stages of the clumps
by comparing the abundances of CCS and HC$_3$N as well as those of CS and SO. 
As the CCS emission has been detected only in dark clouds \citep[e.g.,][]{Nakamura2014}
but not in active cluster-forming clumps \citep[e.g.,][]{Lai2000,Shimoikura2015},
we are particularly interested in carrying out a CCS survey for our targets.

This paper is organized as follows.
In Section \ref{sec:obs}, 
we describe the targets selected for the observations. 
We also describe our observations using the 45 m radio telescope. 
We identify 24 molecular clumps using the \eco\1 data 
and derive their physical properties in Section \ref{sec:results}. 
We analyze the chemical compositions and velocity structures of the clumps,
and discuss how the clumps should evolve in Section \ref{sec:discussion}.
We present the summary of this paper in Section \ref{sec:conclusions}.


\section{OBSERVATIONS} \label{sec:obs}

\subsection{The Target Dust Condensations} \label{sec:target}

\begin{figure*}
\begin{center}
\includegraphics[scale=0.4]{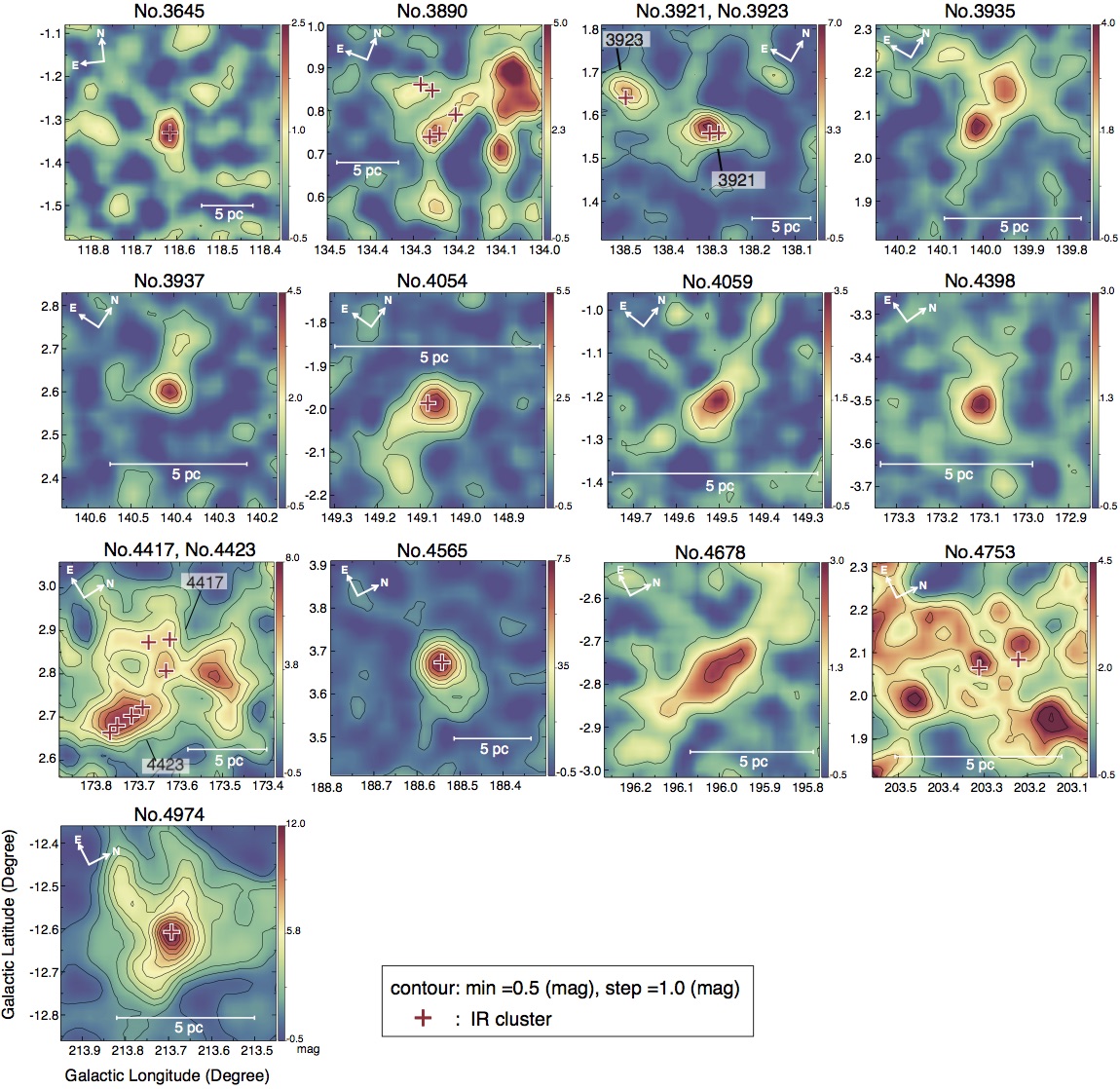}
\caption{
Extinction ($A_V$) maps of the 15 dust condensations generated by \cite{Dobashi2011} 
shown in the Galactic coordinates.
Contours are drawn from 0.5 mag with an increment of 1.0 mag.
Plus signs indicate the positions of the identifed IR clusters
(See Section \ref{sec:cluster}).
\label{fig:coremap}}
\end{center}
\end{figure*}

Based on the catalog of dust condensations of \citet{Dobashi2011} and \citet{Dobashi2013},
we selected 15 dust condensations.
Extinction maps of the 15 dust condensations selected for this study are shown in Figure \ref{fig:coremap}.
Numbers in the figure correspond to the condensation numbers given by \cite{Dobashi2011}.
Except for No. 4678, the dust condensations we selected for the observations are regarded to include IR clusters by \citet[][and assigned Flag `3' in their catalog]{Dobashi2011}, i.e., the dust condensations appear as a `hole' in the $A_{J}$ map due to the increase of star density compared to adjacent regions, but as a `bump' in the $E$($J-H$) map because of the intrinsically red colors of young stars.
We selected the 15 dust condensations from various star forming regions within 2.4 kpc from the Sun.
Except for No. 4398\footnote{
For this region, we calculated the kinematic distance by ourselves 
using the Galactic rotation models in the literature \citep{Kerr,Wouterloot1989,Wouterloot1990}. 
We used the observed \tco peak radial velocity $V_{\rm LSR}=-2.2$ km s$^{-1}$, $R_{0} = 8.5$ kpc, and $\Theta_{0}$ = 220 km s$^{-1}$, 
where $\Theta_{0}$ is the rotational speed of the Galaxy at $R_{0}$.
We obtained a kinematic distance of 0.81 kpc which corresponds to the ``far" distance in the model.
},
the distances to the condensations were obtained from the literature.
We selected relatively massive condensations observable from Nobeyama
with a mass of greater than 100$M_{\sun}$
except for No.4059 ($\sim50$ $M_{\sun}$)
\footnote{
For the mass estimation, we used near-infrared (NIR) color excess map of $E(J-H)$ for the 15 dust condensations from \cite{Dobashi2011}.  
We transformed the extinction values to hydrogen column densities using the relations
N[HI+H$_2]=5.8 \times 10^{21} E(B-V)$ cm$^{-2}$ mag$^{-1}$ \citep{Bohlin} with $R_V=A_V/E(B-V)$=3.1\citep{Cardelli}
and $A_V=E(J-H)/0.107$ \citep{Rieke}.
We then calculated the total mass of the dust condensation as the sum of the extinction values above $A_V=1$ mag.
}.

Ten of the dust condensations are associated with known IR clusters
in active cluster-forming regions such as S201 and Mon R2 \citep[e.g.,][]{Bica_1, Bica_2, Kirsanova2008}.
Typical ages of the IR clusters are $1-5$ Myr.
We also conducted our own analyses using the 2MASS Point Source Catalog (PSC) 
to reveal the star density distribution around the selected condensations, 
which is described in detail in Section \ref{sec:cluster}.
Based on the analyses, we found that the rest of the 5 dust condensations are not associated
with any apparent IR clusters.
The selected dust condensations and their main characteristics are summarized in Table \ref{tab:chara} .

\subsection{Observations with the NRO 45 m Telescope} \label{sec:45obs}

The mapping observations of the 15 dust condensations using the NRO 45 m telescope were carried out during two separate observing periods. 
All of the observations were performed in the on-the-fly (OTF) mode.
The observed molecular lines and the resulting noise levels for each line are summarized in Table \ref{tab:45obs}.

\subsubsection{Observations of $^{12}${\rm CO}, $^{13}${\rm CO} {\rm C}$^{18}${\rm O}, {\rm CS}, {\rm SO} and {\rm C}$^{34}${\rm S}} \label{sec:Tz100}

Observations of the 15 dust condensations with six emission lines at 100 GHz ($^{12}$CO, $^{13}$CO, C$^{18}$O, CS, SO and C$^{34}$S) 
were conducted  for 10 days in February 2013.
The beam size of the telescope is $\sim15\arcsec$ at 110 GHz. 
The single-beam two-polarization two-sideband-separation (2SB) 
sideband-separating Superconductor-Insulator-Superconductor (SIS) receiver TZ \citep{Asayama}
was used as the frontend, and the digital spectrometer SAM45 
with 4096 channels covering a total band width of 31 MHz 
and a velocity resolution of $\sim 0.02$ \kms
was used as the backend.
The overall system temperature ($T_{\rm sys}$) typically ranged from 150 K to 300 K.

During the first part of the observations, the receiver was tuned to the $^{12}$CO \1 line, 
and $\sim 6\arcmin\times6\arcmin$ mapping observations centered on each of the 
dust condensations
were performed along the equatorial coordinates to investigate the molecular distributions.
Afterward, the receiver was tuned to the $^{13}$CO, C$^{18}$O, CS, SO, and C$^{34}$S lines 
to perform simultaneous $\sim 5\arcmin\times5\arcmin$ mapping observations 
centered on the same peaks as those for the $^{12}$CO observations.

An ambient-temperature chopper wheel was employed to determine the antenna temperature scale. 
The telescope pointing for each region was checked every 1.5 hours using SiO maser sources at 43 GHz, 
ensuring that the pointing remained accurate within $\sim 3\arcsec$.
The sideband ratios of the receiver at the rest frequencies of the observed molecular lines were also checked everyday to further calibrate the spectral data, which fluctuated within $\sim10\%$ each time the receiver was tuned.
Calibration was also checked by taking the average spectra of the central $100\arcsec\times100\arcsec$ of the dust condensation No.4423 for all of the molecular lines, confirming consistent calibration within $10\%$ error.

The baselines of the spectral data were removed by fitting the emission-free velocity ranges with linear functions, and the data were converted into three-dimensional data 
with a spheroidal convolution at a angular grid size of $7\farcs5$ 
using the software package {\software{NOSTAR, \citep{Sawada}}.} 
The data were then corrected for the main-beam efficiency of the telescope into the $T_{\rm mb}$ scale
for which main beam efficiencies ($\eta_{\rm mb}$) 
were assumed to be 0.27, 0.30, 0.30, 0.37, 0.38, and 0.39 for the $^{12}$CO, $^{13}$CO, C$^{18}$O, CS, SO and C$^{34}$S lines, respectively. 

We finally obtained the spectral data with an effective angular resolution of approximately $22\arcsec - 24\arcsec$ and a velocity resolution of 0.05 \kms for each emission line. 
The typical 1 $\sigma$ noise level in the $^{12}$CO spectra was $\Delta T_{\rm mb} = 1.2 $ K, and $\Delta T_{\rm mb} =0.2 - 0.5$ K for the other lines.

\subsubsection{Observations of {\rm CCS} and {\rm HC}$_{3}${\rm N}} \label{sec:Z45}

Observations of the 15 dust condensations were also carried out with the CCS and HC$_{3}$N emission lines 
at 45 GHz using the NRO 45 m telescope for three days in March 2017. 
We obtained $\sim 5\arcmin \times 5\arcmin$ maps for the 15 dust condensations in both of the lines simultaneously,
covering the same areas as for the 100 GHz observations.
The dual-polarization receiver Z45 \citep{Nakamura2015} was used,
which provided $T_{\rm sys} \simeq160$ K. 
Spectral data were obtained with the 4096 channels using the SAM45 spectrometer operating at a resolution of 0.1 \kms.
At 45 GHz, the beam size of the telescope was $\sim37\arcsec$ and $\eta_{\rm mb}$ was 0.7.

The telescope pointing and data reduction were performed in the same manner used for the 100 GHz observations.
We smoothed the obtained spectral data to an effective angular and velocity resolutions of  $\sim49\arcsec$ 
and 0.2 \kms, respectively. The resulting 1 $\sigma$ noise level is
$\Delta T_{\rm mb} \simeq 0.2$ K for these resolutions.



\section{RESULTS}\label{sec:results}

\subsection{Identification of Molecular Clumps and Estimation of their Physical Properties} \label{sec:clump}

Figures \ref{fig:iimap1}--\ref{fig:iimap6} show integrated intensity maps of the molecular emission lines 
observed with the 45 m telescope.
In the figures, only molecular lines
significantly detected at the $3\,\sigma$ noise level are shown.
It is seen that the C$^{18}$O line was detected in all of the target regions, 
although some lines such as C$^{34}$S, CCS, and HC$_3$N were not detected
in some of the dust condensations, especially in those without IR clusters (e.g., No. 4398 in Figure \ref{fig:iimap4}).

Based mainly on the C$^{18}$O data, we identified molecular clumps and estimated their physical parameters. 
We used the following four-steps procedure to define the molecular clumps in this study: 

\begin{enumerate}
\item We first searched for pixels having intensities greater than the $3\,\sigma$ noise level in the C$^{18}$O intensity map, and regarded a set of continuous pixels as a clump.
\item We then searched for the peak intensity position in each clump, and defined $S$ the surface area of the clump as the region enclosed by the contour at the $50\%$ of the peak intensity.
\item In the above step 2, we found that four clumps (Nos. 3645, 3890, 4417, and 4565) have two local peaks within $S$ each of which is well separated from the other peak in other molecular lines such as SO and/or CS. 
In addition, the two local peaks have different radial velocities separated by $> 0.5$ km s$^{-1}$. We believe that the two local peaks should originate from two distinct clumps lying along the same line-of-sight. 
We therefore divided each of the four clumps into two smaller clumps at the valley in the C$^{18}$O contours, and re-measured $S$ for the smaller clumps in the same way as described in the step 2.
\item Among the clumps selected by the above steps, we chose only the clumps having $S$ greater than 0.3 arcmin$^2$ (corresponding to 20 pixels in the C$^{18}$O intensity maps) to analyze in this paper in order to ensure the definite detection.
\end{enumerate}

Using the above procedure, we identified 24 clumps in total in the observed 15 dust condensations.
We refer to the clumps as 3645a, 3645b, and so on, in the order of increasing right ascension. 

To examine the star-formation activity in the clumps, 
we investigated the young stellar objects (YSOs) associated with each clump.
The YSOs were searched for by referring to the catalogs of $IRAS$ point sources. 
Eleven clumps were found to be associated with one YSO selected from the $IRAS$ catalog based on the selection criteria proposed by \cite{Dobashi1994}, and they have a bolometric luminosity greater than $10^{2}L_\sun$.

In the following, we estimate physical parameters of the individual clumps. 
To derive column densities from the observed emission lines, we assumed
the local thermodynamic equilibrium (LTE).
Details of the derivations
are described in Appendix \ref{sec:column}.
Based on the \co data,
we calculated excitation temperature $T_{\rm ex}$ at the individual positions of each clump for the observed emission lines. 
Using Equation (\ref{eq:radiative})
and assuming that the \co emission line is optically thick ($\tau\gg1$), $T_{\rm ex}$ can be derived as
\begin{equation}
\label{eq:tex}
T_{\rm ex}=5.53\left\{ \ln\left[1+\frac{5.53}{T^{\rm{co}}_{\rm mb}+0.819}\right]\right\}^{-1} \rm K,
\end{equation}
where $T^{\rm{co}}_{\rm mb}$ is the peak brightness temperature in K of the \co data
estimated by a single Gaussian fit at the individual observed positions. 
The maximum $T_{\rm ex}$ for each clump ranges from 28.0 K (clump 4059) to 82.1 K (clump 4974),
and the average $T_{\rm ex}$ within $S$,  $\overline{T}_{\rm ex}$, ranges from 24.5 K (clump 4059) to 59.4 K (clump 4974). 

Using Equation (\ref{eq:column}) and the derived values of $T_{\rm ex}$, 
we calculated the column densities of \eco, SO, CS, and C$^{34}$S of the clumps
on the assumption of LTE and the optically thin cases ($\tau \ll 1$). 

The molecular hydrogen column density 
$N(\rm H_{2})$ was derived from the \eco column density $N$(C$^{18}$O) 
by assuming the relation $N$(C$^{18}$O)\,$ =1.7\times10^{-7}N(\rm H_{2})$ \citep{Frerking}.
The LTE mass of the clumps within $S$, $M_{\rm LTE}$, was derived as
\begin{equation}
\label{eq:mass}
{M_{\rm LTE}} = \overline{\mu}{m_{\rm H}}\int_S {N({\rm{H_2}})dS},
\end{equation} 
where $m_{\rm H}$ is the hydrogen mass, and $\overline{\mu}$ is the mean molecular weight, 
which is assumed to be 2.8. 
The mean volume density in the clumps, $n(\rm H_{2})$, 
was estimated by dividing $M_{\rm LTE}$ by the clump volume 
V = $(4/3)\pi R^{3}$ where $R$ is the radius of the clump defined as $R=\sqrt{S/\pi}$. 

We also calculated the virial mass, $M_{\rm vir}$, as
\begin{equation}
\label{eq:M_Vir}
M_{\rm vir}=
\frac{5R}{G}
\frac{\overline{\Delta V}({\rm C}^{18}{\rm O})^{2}}{8\,{\rm ln}2}\ ,
\end{equation} 
where $G$ is the gravitational constant, and 
$\overline{\Delta V}({\rm C}^{18}{\rm O})$ is the mean line width (FWHM)
in the composite profile obtained by averaging all of the spectra of the clump within $S$. 
We further derived the average fractional abundances of SO, CS, and C$^{34}$S within $S$ 
by estimating the column densities and dividing them by $N({\rm{H_2}})$ derived from the \eco data.

Using this approach, we found that the clumps have sizes ranging from 0.1 to 0.6 pc, 
masses ranging from 5 to 4397 $M_\sun$, 
and volume densities ranging from $1.2\times10^{4}$ to $1.9\times10^{5}$ cm$^{-3}$.
These physical properties are summarized in Table \ref{tab:mass}.

We note that the $^{12}$CO spectra at some positions around the peak intensities
in the clumps 3890a, 4423a, 4753a, and 4974
are optically thick and suffer from self-absorption. 
For these positions, we cannot determine $T_{\rm ex}$ precisely, which is a source of uncertainty in our mass estimates.
Here we compare our estimates with those in the literature.
\cite{Sakai2006} estimated the LTE mass of clump 3890b (which they called  AFGL333 core A) to be $2300 M_{\sun}$ 
based on the \eco data obtained with the NRO 45 m telescope and assuming a uniform $T_{\rm ex}$ of 20 K.
For the same clump, we estimated $4000 M_{\sun}$ at their assumed distance of 1950 pc, $\sim 70\%$ larger than their value.
As another example, \cite{Peretto} derived a mass of $1650 M_\sun$ for clump 4753b (NGC2264C)
using 1.2 mm continuum data, assuming a dust temperature of 15 K.
Our value for the same clump would be rescaled to $1400 M_\sun$ at their assumed distance of 800 pc, 
which is $\sim 20\%$ smaller than their value.
We found that these differences are mainly due to the difference of the adopted $T_{\rm ex}$ 
and the definition of the surface areas of the clumps. 

We also derived fractional abundances of the CCS and HC$_3$N molecules of the clumps. 
The CCS emission was found only in four clumps (3890a, 4753a, 4753b, and 4974),
whereas the HC$_3$N emission was found in nine clumps (3645b, 3890a, 3890b, 3921, 4423a, 4565b, 4753a, 4753b, and 4974).
All of the clumps detected in CCS were detected also in HC$_3$N.
At an rms noise level of $0.1-0.2$ K and a resolution of 0.2 \kms, there was no detection of both CCS and HC$_3$N
in the other 15 clumps.

For the nine clumps detected in CCS and/or HC$_3$N,
the emission lines were detected only in compact regions in the clumps,
and their peak positions in the integrated intensity maps do not match with those of the C$^{18}$O line.
Therefore, we decided to derive physical properties using these lines at the HC$_3$N intensity peak positions.
To derive the column densities of CCS and HC$_3$N of the clumps, $N$(CCS) and $N$(HC$_3$N),
we tried to estimate $T_{\rm ex}$ by fitting three main hyperfine components of the HC$_3$N spectra.
However, we could not fit the spectra well because of the low signal-to-noise ratios.
We therefore derived $N$(CCS) and $N$(HC$_3$N) using Equation (\ref{eq:column})
on the assumption of a uniform $T_{\rm ex}$ of 5 K measured in dark clouds \citep[e.g.,][]{Hirota2009}
as well as much higher $T_{\rm ex}$ ($\simeq 34-66$ K) measured from the $^{12}$CO spectra, which should give the lower and
upper limits of the column densities, respectively.
We then smoothed the C$^{18}$O spectrum at the HC$_3$N peak position of each clump to $49\arcsec$, 
the same resolution as at 45 GHz,  
and derived $N({\rm{H_2}})$ from  the smoothed C$^{18}$O spectrum in the same manner as
for deriving the clump masses.
Finally, we derived the fractional abundances of CCS and HC$_3$N of the clumps, $f$(CCS) and $f$(HC$_3$N), 
by dividing their column densities by $N({\rm{H_2}})$.
The range of $f$(CCS) in the four clumps lies between $0.5 \times 10^{-10}$ and $4.1 \times 10^{-10}$,
and that of $f$(HC$_3$N) in the nine clumps lies between $0.4 \times 10^{-10}$ and $6.2 \times 10^{-10}$.
For the clumps detected in HC$_3$N but not in CCS, 
we derived the upper limits of $N$(CCS) and $f$(CCS) from the noise levels to be $<1\times 10^{12}$ cm$^{-2}$ 
and $<1\times 10^{-11}$, respectively.
The derived column densities and fractional abundances for the nine clumps are shown in Table \ref{tab:ccs}.

We note that $f$(CCS), $f$(HC$_3$N), $f$(CS), and $f$(SO) derived in the above
may be overestimated, because we derived $N(\rm{H}_2)$ from $N($C$^{18}$O$)$
assuming their flat ratios. In very dense regions,
C$^{18}$O can be less abundant due to adsorption onto dust grains \cite[e.g.,][]{Bergin2002}
or due to selective destruction by the strong UV radiation from young massive stars \cite[e.g.,][]{Shimajiri2014},
which may cause an underestimation of $N(\rm{H}_2)$. 
Taking into account
all of the uncertainties in distance, $T_{\rm ex}$, and $N(\rm{H}_2)$,
we conclude that our estimates of mass and  fractional abundances for the 24 clumps have an uncertainty of 50$\%$ at most.

\subsection{Velocity Dispersion of the Clumps} \label{sec:Velocity}

We investigated the velocity dispersion of the clumps. 
The velocity dispersion ${\sigma_{\rm obs}}$ of the observed emission lines 
at each observed position was derived as
\begin{equation}
\label{eq:sigma}
{\sigma_{\rm obs}}=\sqrt{\frac{\int {(V-{V_0})^{2}}T_{\rm mb}\,dV}{\int T_{\rm mb}\,dV}}\,
\end{equation}
where ${V_0}$ is the intensity-weighted mean velocity measured as ${V_0}=\int VT_{\rm mb}\,dV/\int T_{\rm mb}\,dV$.

Based on the C$^{18}$O data, 
we derived $\sigma_{\rm obs}$ of the emission line, $\sigma_{\rm obs(C^{18}O)}$, from the above equation. 
We regard that $\sigma_{\rm obs(C^{18}O)}$ consists of the thermal (${\sigma_{\rm therm}}$)
and non-thermal (${\sigma_{\rm NT}}$) components \citep[e.g.,][]{Myers}, and can be expressed as
\begin{equation}
\label{eq:sigma}
{\sigma_{\rm obs(C^{18}O)}^{2}}={\sigma_{\rm therm}^{2}}+{\sigma_{\rm NT}^{2}}+{\sigma_{\rm reso}^{2}}\,.
\end{equation}
Here, $\sigma_{\rm reso}$($=0.05$ km s$^{-1}$) is the velocity resolution of the data,
and ${\sigma_{\rm therm}}$ can be calculated as
\begin{equation}
\label{eq:sigma}
{\sigma_{\rm therm}}=\sqrt{\frac{kT_{\rm ex}}{\mu_{(\rm C^{18}O)}m_{\rm H}}}\,
\end{equation}
where $\mu_{(\rm C^{18}O)}$ is the mean molecular weight of C$^{18}$O (=30). 
The three-dimensional velocity dispersion of turbulence at each position can be estimated as
${\sigma_{\rm 3D}}=\sqrt{3} {\sigma_{\rm NT}}$.
The mean values of $\sigma_{\rm obs(C^{18}O)}$, $\sigma_{\rm therm}$, and ${\sigma_{\rm 3D}}$ for each clump
averaged within the clump surface $S$
are listed in Table \ref{tab:disper}.

\subsection{Search for Outflows} \label{sec:outflow}

We investigated the presence of molecular outflows associated with the clumps 
by searching for high-velocity wings in the \co spectra.
The procedure is explained in Appendix \ref{sec:outflow_param}. 
We also searched for molecular outflows associated with the clumps in the literatures \citep[e.g.,][]{Wu}.
As a result, we found nine outflows in the 24 clumps in total, and two of the outflows which are associated with clumps 4417a and 4417c 
are newly identified by this study.
We show the locations of the nine outflows in Figure \ref{fig:outflow}, and summarize their properties in Table \ref{tab:outflow}.

For clump 4753a (NGC2264D), 
\cite{Buckle4753} identified multiple blue and redshifted lobes 
through $^{12}$CO ($J=3-2$) observations.
Our map for this clump does not cover the full extent of the red outflow lobe, and
only the blue lobe is seen in Figure \ref{fig:outflow}. We therefore
list properties of only the blue lobe for this clump in Table \ref{tab:outflow}.


For clump 3921 (AFGL4029), \cite{Snell} found the total outflow lobe mass (the sum of the masses of the red and blue lobes) to be 9.7 $M_{\sun}$ using \co.
Their outflow mass differs significantly from our value of 44.1 $M_{\sun}$ (the lower limit),
which is likely to be explained mainly by their different assumed value of $T_{\rm ex}$ (=25 K). 

We compared the distributions of the molecular outflows with the locations of the IR clusters, and found that 
some of the outflows coincide well with the IR clusters (see Figure \ref{fig:outflow}).
Although the identified outflow lobes of clumps 3890a and 4417a are likely to be unrelated to the corresponding clusters, 
outflow lobes in the other clumps (i.e., clumps 4423a and 4974) coincide well with the positions of the clusters.
In the maps of \cite{Dierickx} produced from their SMA observations, 
clump 4974 (Mon R2) shows the presence of 11 CO outflows 
in the innermost $\sim 1\arcmin$ region around the cluster.
Though only one pair of red and blue lobes is seen in our map,
it is likely that our maps trace the entirety of the high velocity flow originating from the cluster.


\subsection{Search for IR Clusters Based on the 2MASS Data and Estimation of the Physical Properties\label{sec:cluster}}

The 2MASS PSC was used to systematically search for IR clusters associated with all of the 24 observed clumps.
We selected a $15\arcmin\times15\arcmin$ region around each clump except for 
the clump 4974
\footnote{
We investigated how the number of stars in the clusters $N_\star$ of each of our targets changes depending on the survey window size. We investigated $N_\star$ from the star density maps produced for the window size $10\arcmin\times10\arcmin$, $15\arcmin\times15\arcmin$, $20\arcmin\times20\arcmin$, $30\arcmin\times30\arcmin$, $40\arcmin\times40\arcmin$, and $60\arcmin\times60\arcmin$ using the stars except for those with a minor planet flag.
 As a result, we found that the value of $N_\star$ changes depending on the window size only in the case of the cluster associated with the clump 4974. 
For this target, $N_\star$=330 for the size $15\arcmin\times15\arcmin$, but $N_\star$=380 for the size $\gtrsim40\arcmin\times40\arcmin$. 
On the other hand, $N_\star$ of the other clusters do not change for the window size $\gtrsim15\arcmin\times15\arcmin$, because their extents are small.
We therefore selected a $40\arcmin\times40\arcmin$ region around the clump 4974 to produce the star density map.},
and produced star density maps of the regions to identify clusters
using stars which are not recorded as `minor planets' in the 2MASS PSC.
We will explain the reason why we selected the stars in this way in Appendix \ref{sec:cluster_param}.

The star density map of each cluster was constructed by counting the number of stars 
in a $0\farcm4$ radius from each $7\farcs5$ grid set along the equatorial coordinates.
The obtained star density maps still contained contributions from the background sources.
To estimate the properties of the clusters, 
we obtained final star density maps free from the background  
using the $3\,\sigma$ clipping procedure described in detail by \cite{Shimoikura2013}. 
The resulting $1\,\sigma$ noise level of the star density maps is in the range of 3--5 stars arcmin$^{-2}$. 

We searched for pixels in the star density maps with a density greater than the $4\,\sigma$ 
noise level for each region, and regarded them as IR clusters. We then defined the extents of
the clusters by the contour drawn at the $30\%$ level of the peaks.
As a result, we identified 23 IR clusters within the 24 clumps.
Nineteen of the IR clusters have been previously identified in other studies.
In addition to these, we newly identified three IR clusters associated with the dust condensation No. 3890. 
Distribution of the identified IR clusters in the area observed with the 45 m telescope is shown in the last panel of 
Figures \ref{fig:iimap1}--\ref{fig:iimap6}, in which the WISE 3.4 $\micron$ image \citep{WISE} is overlaid
to see the extended emission associated with the clumps or the IR clusters. 
In each panel, the IR clusters are labeled 3645cl, 3890cl1, 3890cl2, etc. 
In summary, the observed regions contain 24 molecular clumps, 
and 16 of them are associated with at least one IR cluster, whereas the other eight clumps 
are associated with no apparent IR cluster. 
Among the 23 clusters identified, six are not associated with any particular clumps. 

To derive properties of the IR clusters,
we first defined their radius as $R_\star=\sqrt{S_\star/\pi}$ 
where $S_\star$ is the surface area of the clusters defined at the $30\%$ contour level. 
We then investigated the cumulative number of stars for each cluster, $N_\star(r)$, 
as a function of distance $r$ from the cluster center. 
An example of the results is shown in Figure \ref{fig:Nstar}.
We found that $N_\star(r)$ increases rapidly with $r$ and becomes rather 
flat at $r \simeq 2R_\star-4R_\star$, and then diverge at larger $r$ due to counting uncertainty,
which is the same trendency that we already found in other cluster-forming regions \citep{Shimoikura2013}. 

We defined the total number of stars contained in an IR cluster, $N_\star$,
as the average value of $N_\star(r)$ in the flattened part of the $N_\star(r)$ vs. $r$ diagram.
For most of the clusters, we defined $N_\star$ as the average value in the range $2R_\star <r< 3R_\star$.
For clusters 3923cl and 4974cl, we decided to define $N_\star$ in the range $4R_\star<r<5R_\star$. 
We found that the $N_\star$ varies in the range of 12 to 380, and $R_\star$ varies from 0.12 to 0.76 pc. 
These parameters are summarized in Table \ref{tab:cluster}.

The largest $N_\star$ ($=380$) in our sample is found in 4974cl which is a rich IR cluster in the Mon R2 region. 
The cluster has already been studied with the 2MASS data by \cite{Carpenter2000} who estimated
the total number of stars to be $N_\star =371$ in an effective radius of $\sim 1.85$ pc based on
the Wainscoat model \citep{Wainscoat}.
We found that their value for $N_\star$ is found to be quite close to ours.

Several studies have used NIR data to derive the structural parameters of clusters 
by fitting the star density profiles with the King's model \citep{King1962}.
Although we attempted to fit the star density of the identified IR clusters with the King model,
most of the clusters could not be fitted well owing to their complex morphologies.
The 23 clusters studied here have a wide variety of structures, and some of them show multiple peaks.
\cite{Camargo2011} studied eight IR clusters among those we study here (clusters 4417cl1$-$cl4 and 4423cl1$-$cl4) 
using the 2MASS PSC and the King model, but they could successfully fit only two of them
(4417cl1 and 4417cl3 that they call  `Sh2-235 Cluster' and `Sh2-235 East2', respectively).

\cite{ladalada} suggested that there are two morphological types of cluster structures, 
hierarchical and concentrated,
and that these structures may reflect the physical processes responsible for cluster formation.
We found that most of our identified clusters are not spherical except for six clusters not associated
with any particular clumps.
For example, cluster 3890cl2, which is located in a cavity of C$^{18}$O gas, has a spherical morphology 
(see Figure \ref{fig:iimap2}).
By contrast, cluster 4423cl2 is elliptical and matches well with its associated clump (see Figure \ref{fig:iimap4}).
We measured the ratio of the minor and major axis with the 2D Gaussian fit for each cluster 
as shown in the last column of Table \ref{tab:cluster},
and found that clusters associated with less molecular gas tend to be more spherical
with a ratio of minor to major axis of $>0.8$,
possibly suggesting that the more spherical clusters are older.

\subsection{Relationships between Clumps and IR Clusters} \label{sec:clump+Cluster}

\begin{figure*}
\begin{center}
\includegraphics[scale=0.4]{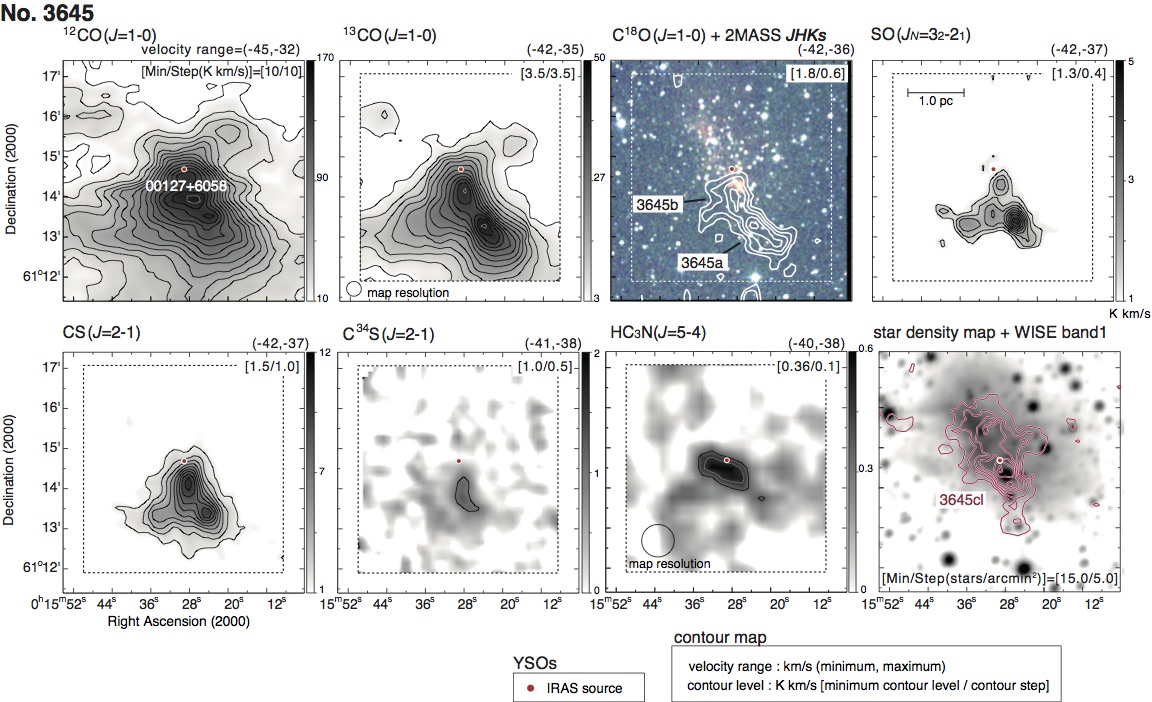}
\caption{
Maps of the integrated intensity of the observed molecular lines around the dust condensation No.3645 
obtained with the 45 m telescope, 
shown in the equatorial coordinates.
Molecules, transitions, and the velocity ranges in units of \kms used for the integration are indicated above the maps. 
The lowest contours and the contour intervals in units of K \kms are indicated in the square brackets in each panel.
No obvious CCS emission is detected in this dust condensation.
The map resolution is indicated by the open circle in panels for the $^{13}$CO and HC$_3$N emission lines.
The last panel on the lower right shows the star density map of the cluster 3645cl 
overlaid with the WISE 3.4 $\mu$m image (gray scale). 
The lowest contour and the contour interval in units of arcmin$^{-2}$ are indicated in the panel. 
The red dot denotes the position of the selected $IRAS$ source.
\label{fig:iimap1}}
\end{center} 
\end{figure*}

\begin{figure*}
\begin{center}
\includegraphics[scale=0.4]{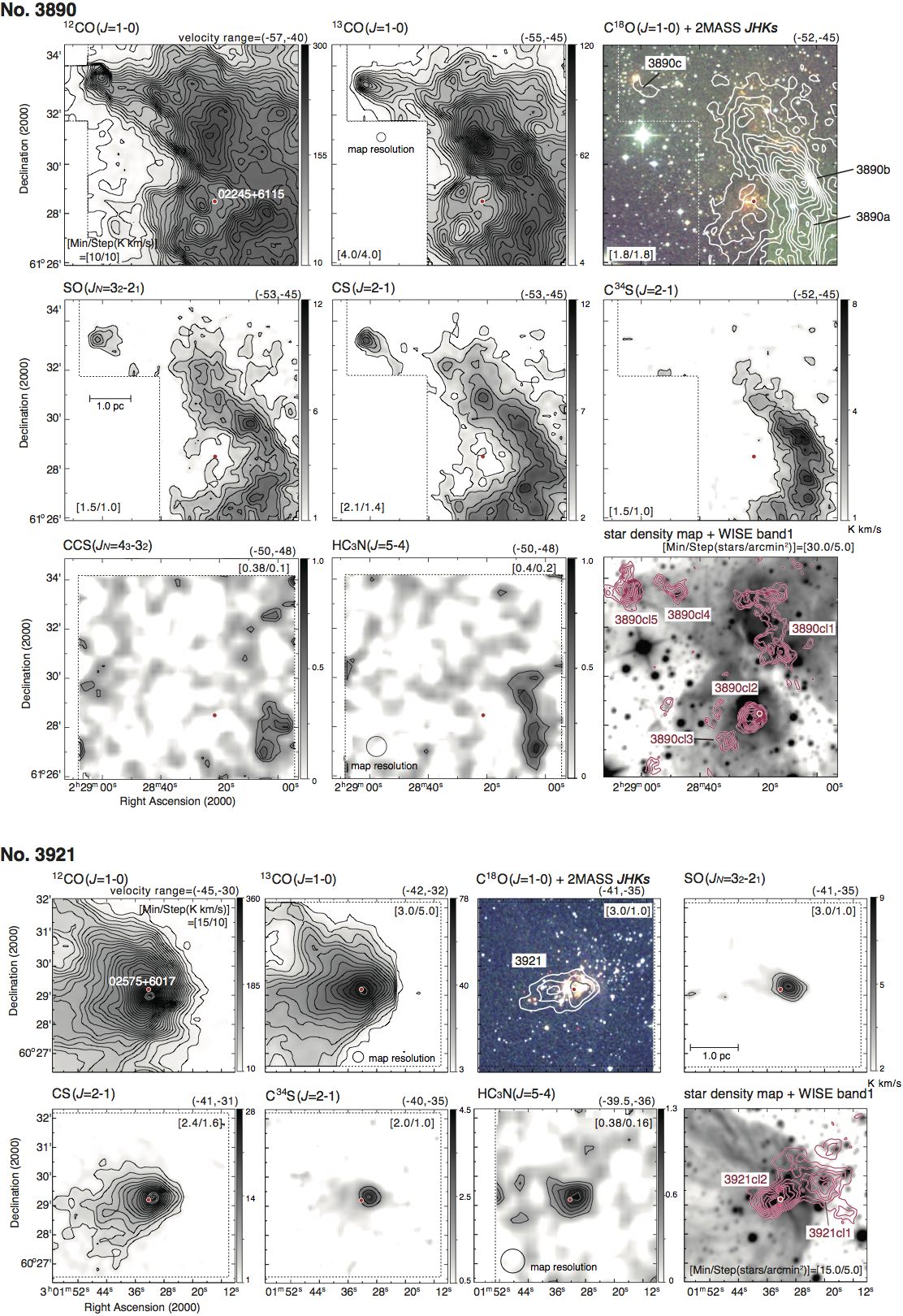}
\caption{
Same as Figure \ref{fig:iimap1}, but for the dust condensations Nos. 3890 and 3921. 
\label{fig:iimap2}}
\end{center} 
\end{figure*}

\begin{figure*}
\begin{center}
\includegraphics[scale=0.4]{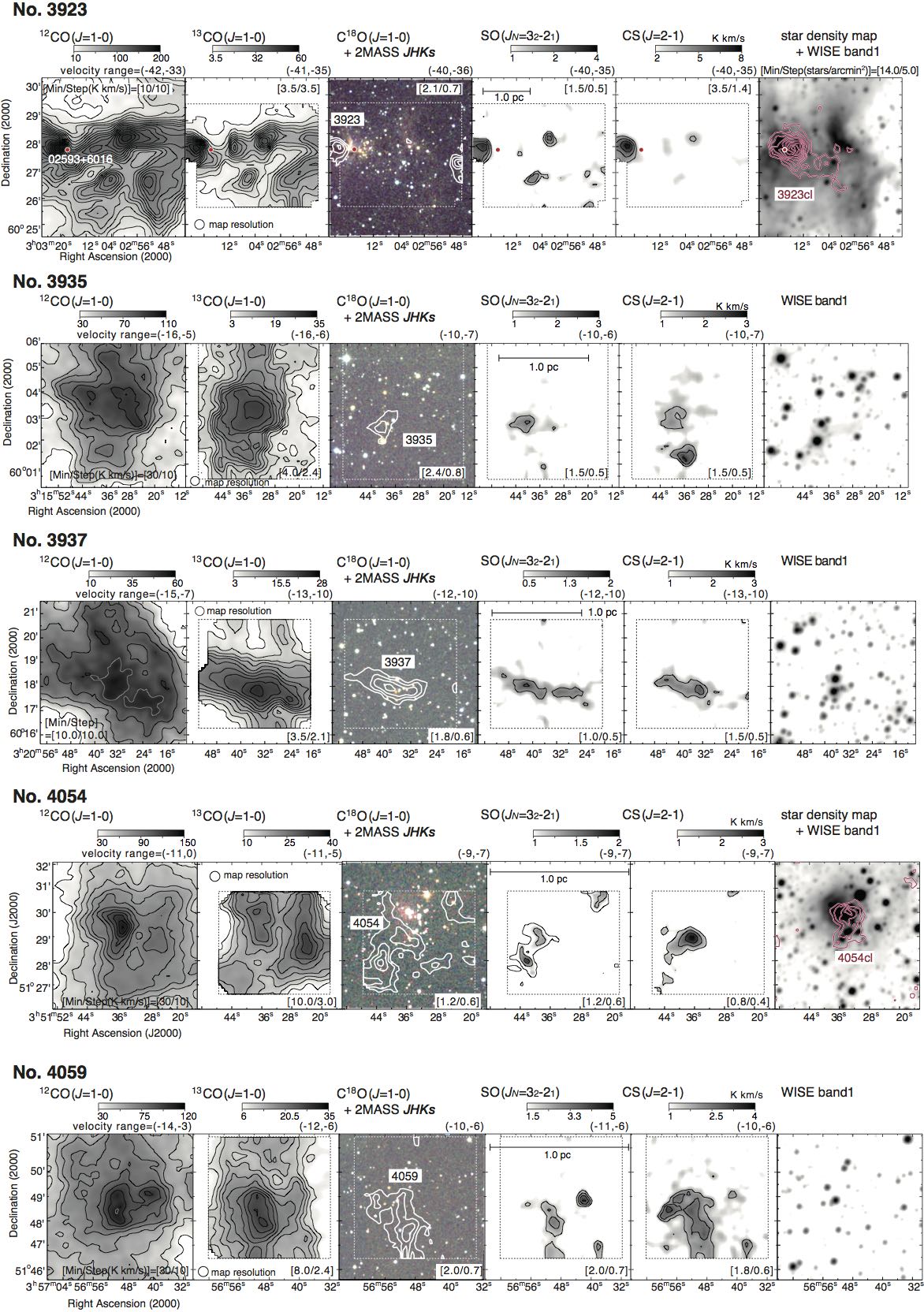}
\caption{
Same as Figure \ref{fig:iimap1}, but for the dust condensations Nos. 3923, 3935, 3937, 4054, and 4059. 
No obvious C$^{34}$S emission is detected in all of the dust condensations.
No obvious IR cluster is seen in the dust condensations Nos. 3935, 3937, and 4059.
\label{fig:iimap3}}
\end{center} 
\end{figure*}

\begin{figure*}
\begin{center}
\includegraphics[scale=0.4]{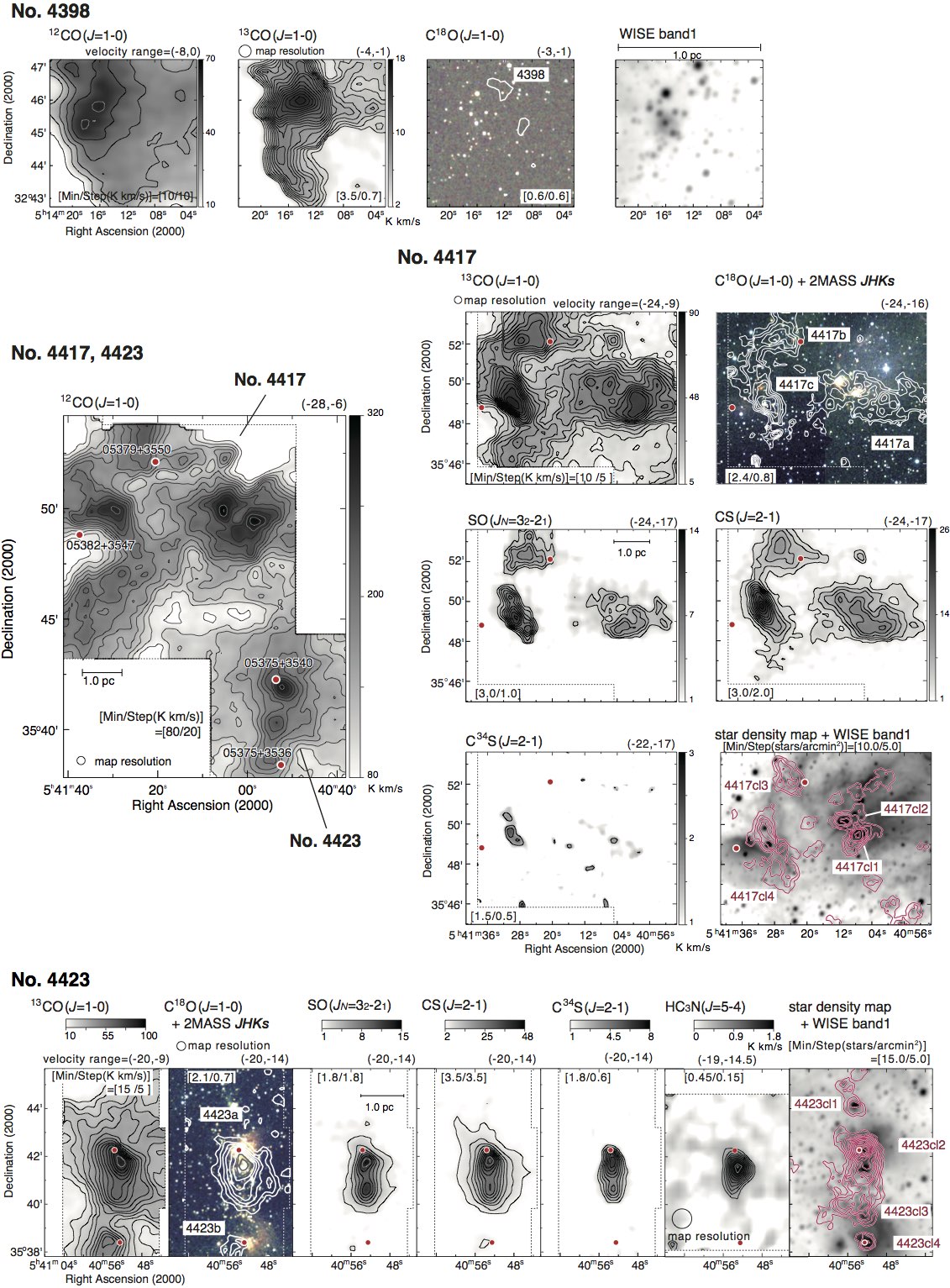}
\caption{
Same as Figure \ref{fig:iimap1}, but for the dust condensations Nos. 4398, 4417, and 4423. 
No obvious SO, CS, C$^{34}$S emission or IR cluster are seen in the dust condensation No.4398. 
\label{fig:iimap4}}
\end{center} 
\end{figure*}

\begin{figure*}
\begin{center}
\includegraphics[scale=0.4]{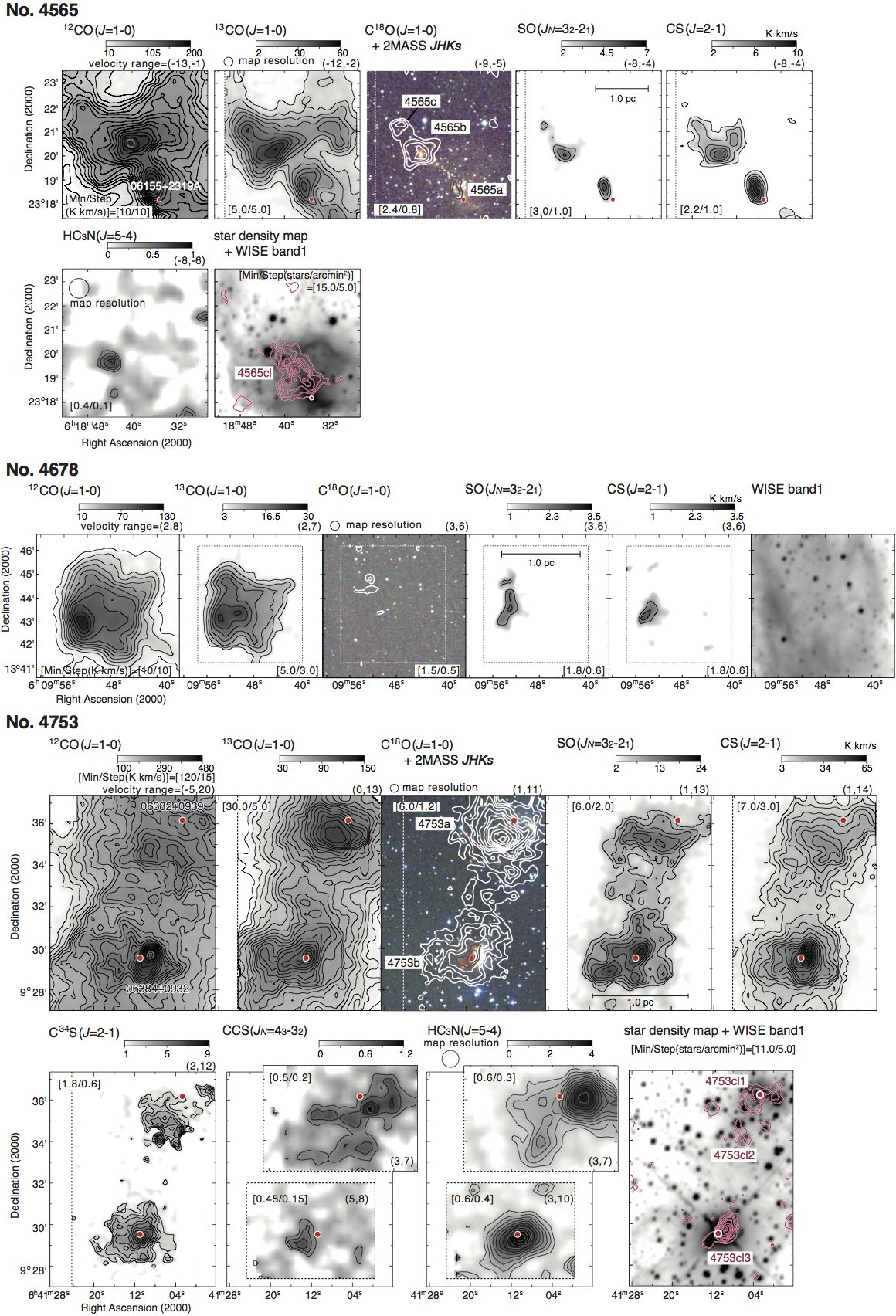}
\caption{
Same as Figure \ref{fig:iimap1}, but for the dust condensations Nos. 4565, 4678, and 4753. 
No obvious C$^{34}$S emission is detected in the dust condensations Nos. 4565 and 4678. 
No obvious IR cluster is seen in the dust condensation No.4678.
\label{fig:iimap5}}
\end{center} 
\end{figure*}

\begin{figure*}
\begin{center}
\includegraphics[scale=0.4]{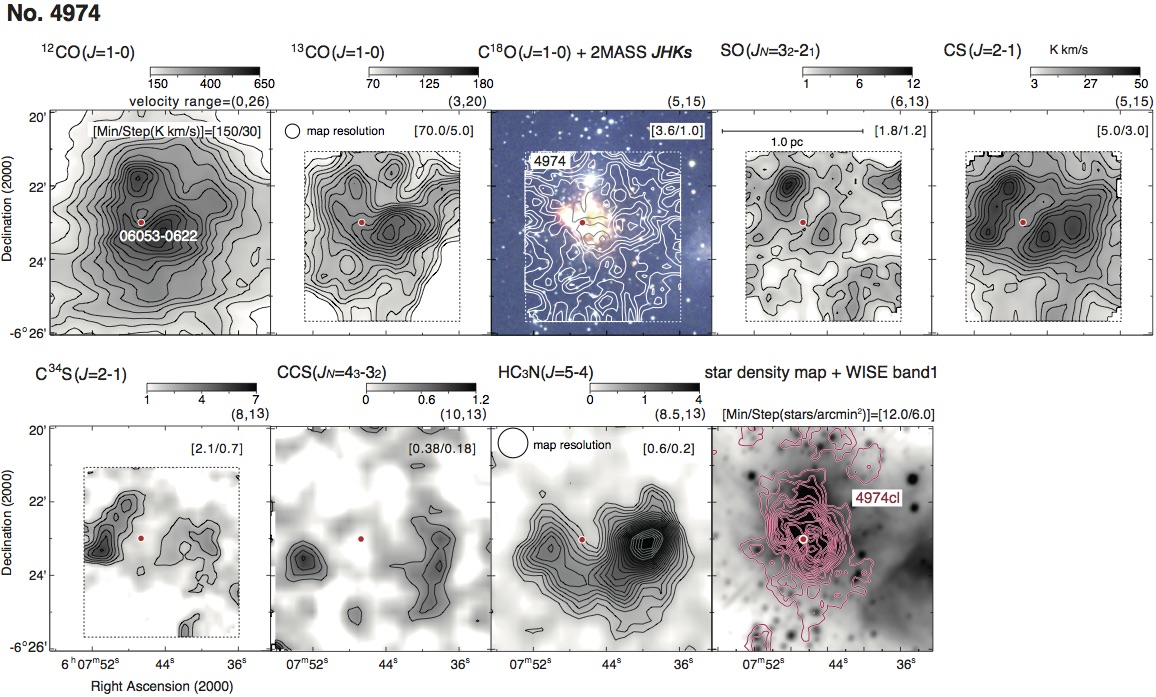}
\caption{
Same as Figure \ref{fig:iimap1}, but for the dust condensation No. 4974. 
\label{fig:iimap6}}
\end{center} 
\end{figure*}

By comparing the distributions of the clumps with those of the associated IR clusters (Figures \ref{fig:iimap1}--\ref{fig:iimap6}), 
we found that their spatial coincidences vary from clump to clump.
For example, in some cases, there are two clumps within the extent of one cluster (e.g., clusters 3645cl and 4565cl), 
while there are sometimes two IR clusters within the extent of one clump (e.g., clump 4753a).
We also note that the IR clusters in clump 3921 can be resolved by eye into two IR clusters, 3921cl1 and 3921cl2, 
and the former cluster is not associated with any molecular emission. 

There have been some studies to investigate evolution of cluster forming clumps,
e.g., by \citet{Ridge2003}, \citet{Kawamura2009}, \citet{ Higuchi2009}, and \citet{Shimoikura2013}.
In their studies, clumps are classified by eye inspection mainly according to the
presence of the associated clusters and their morphological relations.
To examine the relationship between the clumps and the IR clusters
and to classify their morphological types more quantitatively,
we investigated the spatial correlations between $N$(\eco)  
and the star densities (SD) of the associated clusters
for the 16 clumps.
To do this, we re-gridded and smoothed the data sets to the same angular resolution ($48\arcsec$)
and performed a linear least-square fit to quantify the $N(\rm {H}_2)$ vs. SD relation. 
The results are shown in Figure \ref{fig:SD_NH2}.
Correlation coefficients of the fit, $r$, ranges from 0.1 to 0.8. 

To characterize the identified clumps and clusters in terms of the cluster formation, 
we classified them into four types according to the following correlation relationship:

\begin{enumerate}
\item We classify clumps associated with no apparent IR cluster as Type 1.
\item We classify clumps that are well-correlated with IR clusters 
with a correlation coefficient of $\gtrsim0.5$ as Type 2. 
Such clumps have similar morphologies to the clusters.
\item We classify clumps not well-correlated with IR clusters with a correlation coefficient of $<0.5$ as Type 3. 
In such clumps, the peaks of the star density of the IR clusters do not match with those of the clumps.  
\item Finally, 
we classify clusters not associated with any clumps as Type 4.
\end{enumerate}

Among the 24 clumps, eight, seven, and nine clumps are classified as Type 1, 2, and 3, respectively.
In addition, six IR clusters are classified as Type 4.
The classification of the clumps and IR clusters are summarized in Table \ref{tab:clump+cluster}.


\subsection{Physical Properties of the Clumps} \label{sec:corr}

We investigated the correlations between the physical properties of the clumps 
using the derived parameters listed in Tables \ref{tab:mass}, \ref{tab:disper}, and \ref{tab:outflow}. 

Figure \ref{fig:corr1} shows the relations among $M_{\rm vir}$, $M_{\rm LTE}$, $R$, and $\Delta V$ 
for the clumps of Types 1, 2, and 3, which are shown in different colors.
We also plotted the same relations for clumps identified in the S252, S247, 
and BFS52 regions using \eco data presented by \cite{Shimoikura2013}, which
were also obtained with the NRO 45 m telescope.
We classified these clumps into the three types (Types 1--3) in the same way as explained above.

As shown in Figure \ref{fig:corr1}(a), many of the clumps have a mass of the same order as their virial masses,
suggesting that they may be in the virial equilibrium.
However, there is a tendency that masses of the massive clumps ($M_{\rm LTE} \gtrsim10^3$ $M_\sun$)
are significantly higher than the virial masses. These clumps can be collapsing, if they are not supported by
the clump-supporting forces such as the magnetic filed. In fact, as we will show in Section 4.2,
some of the Type 2 clumps are actually collapsing.

The relationship between $M_{\rm LTE}$ and $R$ in Figure \ref{fig:corr1}(b) 
reveals that the clumps have a molecular number density in the range 10$^{4} < n(\rm H_2)< 10^{5}$ cm$^{-3}$,
which is consistent with the results of previous studies of cluster-forming clumps \cite[e.g.,][]{Saito2007,Shimoikura2013}.

In Figure \ref{fig:corr1}(c), it is apparent that the clumps with clusters, 
particularly Type 2 clumps, have larger line widths than those without clusters. 
We also found that the clumps having large line widths are associated with outflows,
suggesting that the outflows are supplying turbulence to the clumps.

It has been long suggested that outflows are an important source of turbulence of
natal clumps \citep[e.g.,][]{Snell,Nakamura2011}.
In order to see how much the outflows can contribute to the observed turbulence,
we estimated the dissipation rates of turbulence of the clumps as
$\dot{P}_{\rm turb}=-0.21M_{\rm LTE}\,{\sigma_{\rm3D}^2}/R$ \citep[][see their equations 4 and 5]{Nakamura2014a} using parameters in Tables \ref{tab:mass} and \ref{tab:disper},
and compared them with $\dot{P}_{\rm flow}$ the ejection rates of momentum of the outflows in Table \ref{tab:outflow}.
Results are shown in Figure \ref{fig:Pdot}.
For $\dot{P}_{\rm flow}$, we plotted the geometrical mean values with error bars representing the minimum and maximum estimates (see Appendix \ref{sec:outflow_param}).
As seen in the figure,
$\dot{P}_{\rm flow}$ is comparable to or larger than $\dot{P}_{\rm turb}$ for all of the Type 2 and 3 clumps within the errors, 
suggesting that the observed turbulence can be maintained by the outflows for these clumps.
For the Type 1 clump (3890a), however, $\dot{P}_{\rm flow}$ is significantly smaller than $\dot{P}_{\rm turb}$,
and thus the observed turbulence of this clump should dissipate
on a scale of crossing time ($R/\sigma_{\rm 3D}\simeq 3.8\times 10^5$ yr),
if the clump does not form additional outflows in the future.

Figure \ref{fig:corr1} suggest that all of the Type 2 clumps have similar properties with a 
typical mass and radius of $\sim10^{3} M{_\sun}$ and 0.5 pc, respectively. 
By contrast, the properties of the other type clumps are widely distributed in the diagrams. 
Two Type 1 clumps, 3890a and 3890b, have a very large mass of $\sim 4000 M{_\sun}$
and they are lying in a locus similar to that of the Type 2 clumps in Figure \ref{fig:corr1},
suggesting that they are potential sites of cluster formation in the future.
However, as we will show in Section \ref{sec:composition}, the Type 1 clumps exhibit
puzzling chemical compositions.
We will discuss the origin of the Type 1 clumps in Section \ref{sec:evolution}.

\section{DISCUSSION} \label{sec:discussion}

At a glance, we would expect that the clump + cluster systems
should evolve from Type 1 to Type 4 successively, but this needs a careful inspection
of the obtained data set.
In the following, we will examine the chemical compositions of the clumps
which can be used as a measure of their ages in terms of chemical reaction.
We will then inspect the velocity structures as well, and will discuss how the
the clump + cluster systems should evolve.


\subsection{Chemical Compositions of the Clumps}\label{sec:composition}

Up to date, the CCS emission has been detected mostly toward
dense regions in low-mass star forming regions, but not in active
cluster forming regions such as DR21, W3, or W40 \citep[e.g.,][]{Lai2000,Shimoikura2015}.
In this study, we detected the CCS emission for the first time in active cluster forming clumps,
i.e., toward the three Type 2 clumps 4753a, 4753b, and 4974. The CCS emission was also detected
in one non-cluster forming Type 1 clump (3890a), which was already reported by \cite{Sakai2006}.

The relationships between the clumps and the IR clusters suggest that the clumps classified 
as Types 1 -- 3 are in different stages of evolution. 
To investigate this in terms of chemical reactions, 
we compared $f$(CCS) and $f$(HC$_3$N) for the nine clumps in which HC$_3$N was detected.
The relation is shown in Figure \ref{fig:CCS}.
As explained in Section \ref{sec:clump}, we estimated the upper and lower limits of the
fractional abundances for different assumption of $T_{\rm ex}$.
We plotted their geometrical mean values with error bars representing the two limits.
In the figure, we also show results of calculations based on a chemical model by \cite{Suzuki1992}. 
They carried out single-dish survey observations toward 49 dark cloud cores in low-mass star forming regions,
and also performed the model calculations to find 
that CCS and HC${_3}$N are more abundant in an early stage of chemical evolution. 
Their results were also confirmed by \cite{Hirota2009}. 
The fractional abundances of CCS and HC${_3}$N 
can therefore be regarded as a good measure of chemical reaction time in
dense cores just before/after the formation of stars therein.

In Figure \ref{fig:CCS}, we also show the $f$(CCS) and $f$(HC$_3$N) data points for dark cloud cores
for comparison, which we reported earlier \citep{Shimoikura2012}.
We note that the plots for the Types 2 and Type 3 clumps in the figure
are likely to be shifted from the global trend of the dark cloud cores,
though the ambiguity is large.
Although we do not understand clearly the reason for the shift at the moment, it may reflect
a significant difference in chemical reactions in massive clumps with IR clusters and small dark cloud cores.

To better establish the evolutionary stages of the Type 1$-$3 clumps,
we compared in Figure \ref{fig:SO_CS} $\overline{f}$(CS) the average fractional abundance of CS
and $\overline{f}$(SO) that of SO
within the clump surface $S$ for the 24 clumps.
We also plot the same relation for the clumps in our earlier paper \citep{Shimoikura2013}. 
It is also known that CS and SO can be used as a chemical clock \citep[e.g.,][]{Bergin}, and
it is expected that CS should appear in an early stages of chemical reactions in molecular clumps, 
while SO should appear in a later stage.
We also show a theoretical model calculated by \cite{Bergin} in the figure. 
According to the calculations based on their ``model 1", $f$(CS) and $f$(SO) start to decrease monotonically at 
$\sim2\times10^{5}$ yr and $\sim2\times10^{6}$ yr, respectively.

As seen in Figure \ref{fig:SO_CS}, the distribution of the clumps can be well separated according to their types.
The figure shows that $\overline{f}$(CS) and $\overline{f}$(SO) are decreasing
in the order of Type 2 , Type 3, and Type 1.
In general, difference in the chemical composition can be caused not only by
the ages but also by the density of the clumps. However, the latter 
may not be an important cause for the clumps studied here,
because most of them have similar densities in the range $10^4-10^5$ cm$^{-3}$
(Figure \ref{fig:corr1}).

We further compared the
$f$(CCS) vs. $f$(HC$_3$N) and $\overline{f}$(CS) vs. $\overline{f}$(SO) relations with known cluster ages.
The ages of the clusters associated with the clumps 4753a and 4753b 
are estimated to be $\sim3$ Myr \citep[for NGC2264, e.g.,][]{Sung}
and those of clump 4974 is estimated to be $\sim1-4$ Myr \citep[for Mon R2, e.g.,][]{Maaskant}.
As seen in Figure \ref{fig:CCS} and \ref{fig:SO_CS}, it is interesting to note that
the ages of the clusters associated with the Type 2 clumps ($\sim 1-4$ Myr) are consistent
with the model calculations by \cite{Suzuki1992} and \cite{Bergin}.
However, we should note that the models 
assume the gas temperature of only 10 K for dark clouds
and do not take into account the feedback of star formation,
and thus we cannot directly compare the ages of the clusters with
the chemical reaction time of the models.
Though it is difficult to take into account the feedback of star formation,
recent chemical model calculations for much warmer clumps \citep[up to 200 K, e.g.,][]{Chapman,Hassel}
show that fractional abundances of carbon-chain species such as CCS and HC$_3$N
generally decrease in the time range roughly from $\sim 1\times 10^5$ yr to $\sim1 \times 10^6$ yr
in the same way as for cold dark clouds (10 K) but with shorter reaction time for higher gas temperatures.
Because the observed clumps are warmer in the order of the Type 2,
Type 3, and Type 1 clumps (see Tables \ref{tab:mass} and \ref{tab:ccs}),
the loci of the Types 1--3 clumps in Figures
\ref{fig:CCS} and \ref{fig:SO_CS} indicate that the Type 2 clumps are the youngest and the Type 1 clumps
are the oldest among the three types in terms of chemical compositions.

 
In summary, the differences in the fractional abundances of
CCS, HC$_3$N, CS, and SO among the three types of the clumps 
are likely to reflect the different evolutionary stages of the clump types:
The Type 1 clumps are seen in the lower-left side of the diagram in Figure \ref{fig:SO_CS}, 
suggesting that they may be rather old in terms of chemical compositions. 
The Type 2 clumps are in an early stage of cluster formation,
in which the associated clusters are still embedded in an amount of gas and dust.
The Type 3 clumps are in a more evolved stage than the Type 2 clumps with much less molecular gas.
The Type 4 clusters with no or very little gas should be in the final stage of cluster formation.
We suggest that the massive clumps should evolve in the order
of Type 2, 3, and 4 successively.

The Type 1 clumps are puzzling, because we would expect that they are the clumps 
just before the onset of cluster formation in a stage prior to the Type 2 clumps, but
their chemical compositions indicate that they are older than the Type 2 clumps.
We will discuss the nature and origin of the Type 1 clumps in Section \ref{sec:evolution}.



\subsection{Velocity Structures of the Clumps}

The results shown in Figures \ref{fig:corr1}-- \ref{fig:Pdot}
suggest that there are interesting systematic differences among the types of the clumps.
In the following, we further investigate the velocity structures of the Types 1--3 clumps. 

Figure \ref{fig:sigma} shows the distribution of the velocity dispersion ${\sigma_{\rm obs}}$ 
measured with the \tco and \eco emission lines for the 24 clumps
as a function of the projected distance from the cluster center (for Type 2 and Type 3 clumps) 
or from the clump peak (for Type 1 clumps). 
Values of ${\sigma_{\rm obs}}$ measured with the two emission lines are shown by dots with different colors;
gray and pink for the \tco and \eco emission lines.
We found that ${\sigma_{\rm obs}}$ at the center of the clumps differs depending on the types of the clumps;
the ${\sigma_{\rm obs}}$ values of Type 1 (except for clumps 3890a and 3890b), Type 2 (except for 4417b), and Type 3
are mostly $\lesssim 0.5$ \kms, $\gtrsim 1.0$ \kms, and $\gtrsim 0.5$ \kms, respectively.
This suggests that the large velocity dispersion at the clump center is
a typical feature of cluster-forming Type 2 and Type 3 clumps.
The velocity dispersion is especially large in the Type 2 clumps, indicating that there must be
a lot of motions around the clump center in the beginning of cluster formation.

We also investigated the \eco distributions based on the position-velocity (PV) diagrams 
for all of the clumps. In Figure \ref{fig:PV}, we show examples of the PV diagrams
measured along the major and minor axes of the clumps, which are displayed in 
panels (b) and (c) of the figure.
As denoted by the white broken circles in panels (b),
we found that the PV diagrams
of some Type 2 clumps
measured along the major axis
show double peaks separated by $\sim1$ \kms, and they are
symmetrically located with respect to the clump centers.
This feature is seen only in three of the seven Type 2 clumps (4417c, 4423a, and 4753b),
but not in Type 1 or Type 3 clumps.

A possible interpretation for the double-peaked feature of the Type 2 clumps
could be a collision of two smaller clumps \cite[e.g.,][]{Higuchi2009}, but in that case,
we would expect that the velocity difference of the two peaks should
largely vary, because the collision velocity is a free parameter and can take any values basically.
However, the velocity differences seen in the PV diagrams for clumps 4417c, 4423a, and 4753b
are rather small being always only $\lesssim 1$ \kms,
suggesting that the double-peaked feature in the PV diagrams is unlikely to be caused
by the clump collisions.

To understand the velocity structure of these Type 2 clumps,
we analyzed the C$^{18}$O data of clump 4423a (S235AB)
in our previous paper \citep{Shimoikura2016}, which shows
the double-peaked feature most clearly among the three Type 2 clumps,
and concluded that the observed feature around the clump center
should be caused by the infalling motion with rotation
of the entire clump, not by the collisions of smaller clumps.
Note that the double-peaked feature in the PV diagram is very
similar to those observed around dense cores forming a single low-mass YSO, 
which was successfully interpreted by \cite{Ohashi1997} as an infalling core with rotation. 
As we described in details in the previous paper \citep{Shimoikura2016},
we made a simple model of a massive cluster forming clump ($\sim 1000 M_\sun$)
with analog to the model suggested by \cite{Ohashi1997} for a small core forming a single YSO
(see their Figure 8 and Appendix), 
and we confirmed that the model can reproduce well the PV diagrams observed toward clump 4423a.
Obviously, the other Type 2 clumps shown in Figure \ref{fig:PV} (4417c and 4753b) also
exhibit the double peaks in the PV diagram, indicating that these clumps should also be infalling,
although they are too complex to fit nicely with our simple model.
Actually, the collapsing motion of clump 4753b was suggested also by \cite{Peretto}
who found the two velocity components through molecular line observations 
using the IRAM 30 m telescope.

The above findings imply that, in the beginning of cluster formation,
the massive cluster-forming clumps ($\sim1000 M_\sun$)
have a simple structure similar to a small dense core forming a single
low mass star, and we would expect
that the dense massive core and therefore the most massive star(s) in the cluster
should form at the center of the clump. Such a scenario is supported
by recent numerical simulations \citep{Smith2009,Wang2010}.
In the case of the three Type 2 clumps shown in Figure \ref{fig:PV}
(4423a, 4417c, 4753b),
some observational evidences for the formation of massive cores and
YSOs at the clump center can also be found in the literature
\citep{Felli2004, Peretto, Dewangan2011}.

The fact that at least three of the seven Type 2 clumps exhibit the double-peaked feature
indicates that the infalling motion with rotation of the entire massive clumps is a common
phenomenon in the Type 2 clumps.
We note that all of
the Type 2 clumps showing the highest $f$(CS) and $f$(SO) are the
ones exhibiting the double-peaked feature in the PV diagrams,
indicating that the dynamical infall with rotation occurs in the beginning
of cluster formation.

For the rest of Type 2 clumps (3921, 4417b, 4753a, 4974),
the double-peaked feature is not seen in our data set, but
it can be due to rather poor angular resolution of the observations (22$\arcsec$).
In fact, the clumps show large velocity gradient along their major axes
similar to those observed toward the three clumps with the double-peaked feature,
suggesting that they are collapsing but the double-peaked features are not
resolved by the current observations.



\subsection{Nature of the Observed Massive Type 1 Clumps and the Possible Evolutionary Scenario of the Cluster Formation}\label{sec:evolution}

As discussed in Section 4.1, 
the observed Type 1 clumps are generally older than the Type 2 clumps in terms of chemical compositions.
Some of the small Type 1 clumps ($\sim100$ $M_\sun$) can be remnants of the Type 3 clumps,
or they might be too small to produce clusters and have survived for a long time without collapsing
for some reasons, e.g., by maintaining turbulence by outflows driven by one or a few low-mass YSOs forming there 
\citep{Snell1987,Dobashi1993,Nakamura2011},
or by the support of the magnetic field like what has been suggested for small clouds in the Ophiuchus north region \citep{Nozawa1991,Hirota2009}.
However, the massive Type 1 clumps, namely 3890a and 3890b with a mass of $\sim 4000$ $M_\sun$
are rather mysterious, because, except for their puzzling chemical compositions, they appear to be
the very initial clumps that should evolve to Type 2 clumps in terms of the mass and velocity structures.
Here, we focus on the nature of these massive Type 1 clumps.

\begin{figure*}
\begin{center}
\includegraphics[scale=0.4]{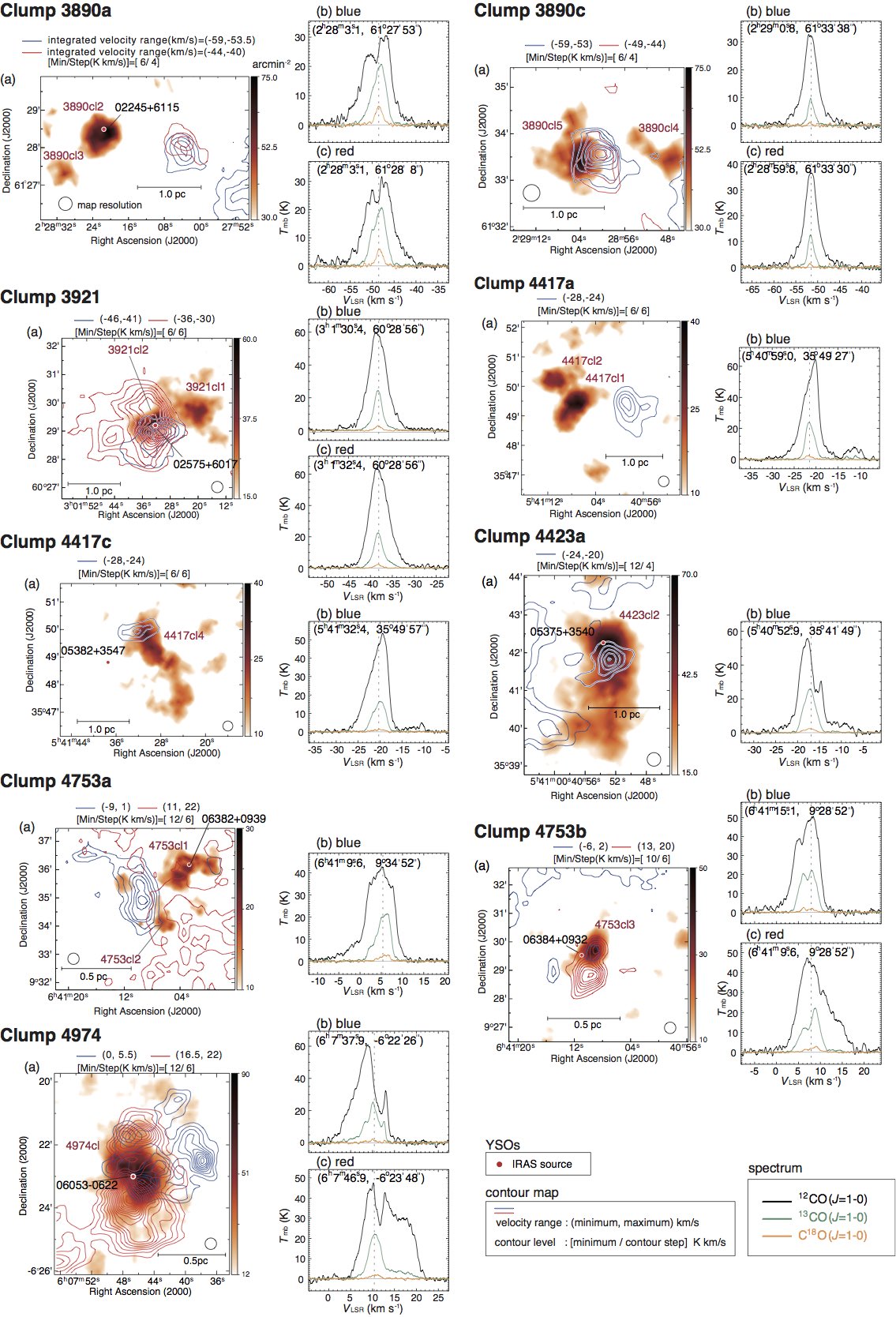}
\caption{
The $^{12}$CO intensity maps for the blue and/or red lobes of the molecular outflows
found around the clumps are shown in panel (a). The red scale denotes the star density distributions. 
The red contours denote the red lobes, and the blue contours denote the blue lobes.
The contours start from the 3 $\sigma$ noise level.
Sample spectra obtained at the intensity-peaks of 
the blue lobes and the red lobes are shown in panels (b) and (c), respectively. The equatorial coordinates are given in each panel. 
The vertical broken line in the spectra marks the systemic velocity.
\label{fig:outflow}}
\end{center} 
\end{figure*}

\begin{figure*}
\begin{center}
\includegraphics[scale=0.4]{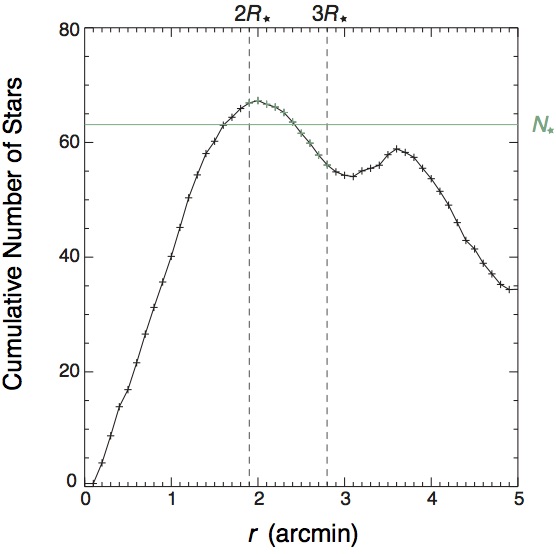}
\caption{Cumulative number of stars as a function of radius for the cluster 3645cl. The number of stars
were measured with the background-subtructed star density map.
The number of stars of the cluster $N_\star$ (indicated by the green line) was defined as
the mean number of stars at  $r=2R_\star$ and $3R_\star$ where $R_\star$ is the cluster radius (see text).
\label{fig:Nstar}
}
\end{center} 
\end{figure*}

\begin{figure*}
\begin{center}
\includegraphics[scale=0.4]{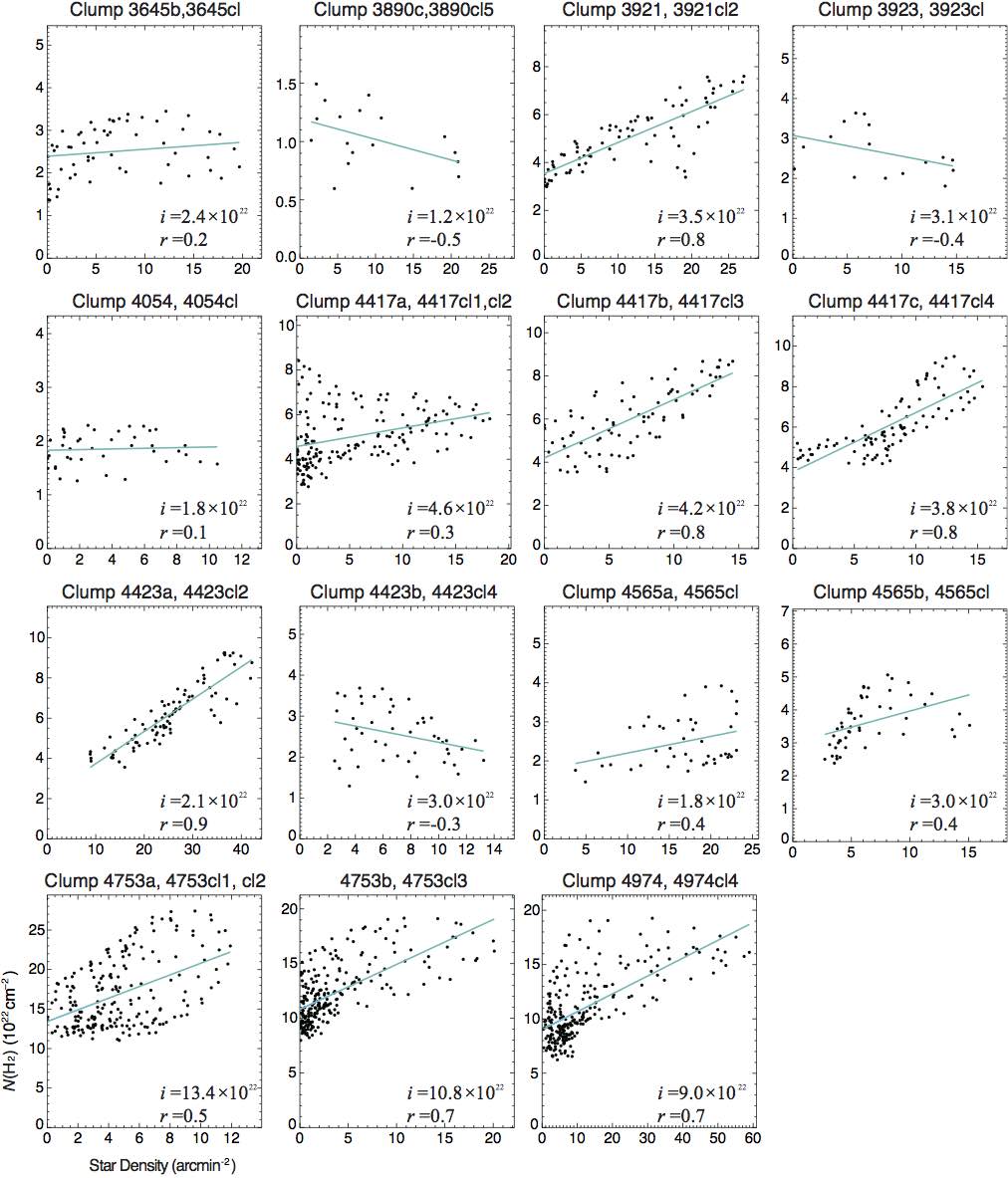}
\caption{
Correlation between $N(\rm H_2)$ and the star density of the 16 clumps with IR clusters.
The blue lines denote the linear least-square fit.
The intercepts $i$ and correlation coefficients $r$ for the relations are denoted in the panels.
\label{fig:SD_NH2}}
\end{center} 
\end{figure*}

\begin{figure*}
\begin{center}
\includegraphics[scale=0.4]{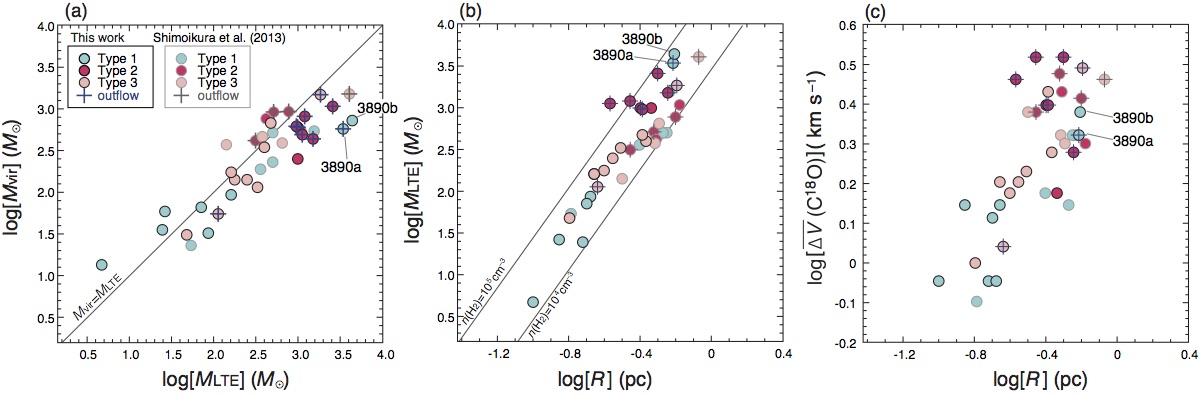}
\caption{
Relation between 
(a) $M_{\rm vir}$ and $M_{\rm LTE}$, (b) $M_{\rm LTE}$ and $R$, 
and (c) $\overline{\Delta V}$(C$^{18}$O) and $R$ for the 24 clumps.
Blue, red, pink circles represent the clumps of the different types.
Plus signs denote the clumps associated with a molecular outflow. 
We also show data points of other clumps presented by \cite{Shimoikura2013}.
\label{fig:corr1}}
\end{center} 
\end{figure*}

\begin{figure*}
\begin{center}
\includegraphics[scale=0.5]{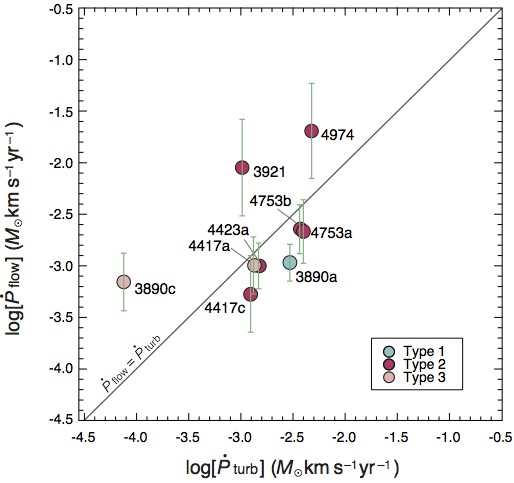}
\caption{
The $\dot{P}_{\rm flow}$ vs. $\dot{P}_{\rm turb}$ relation for the clumps associated with outflows.
For $\dot{P}_{\rm flow}$, we plot the geometrical mean values 
with error bars (green line) representing the minimum and maximum estimates (see Appendix \ref{sec:outflow_param}).
\label{fig:Pdot}}
\end{center} 
\end{figure*}

\begin{figure*}
\begin{center}
\includegraphics[scale=0.5]{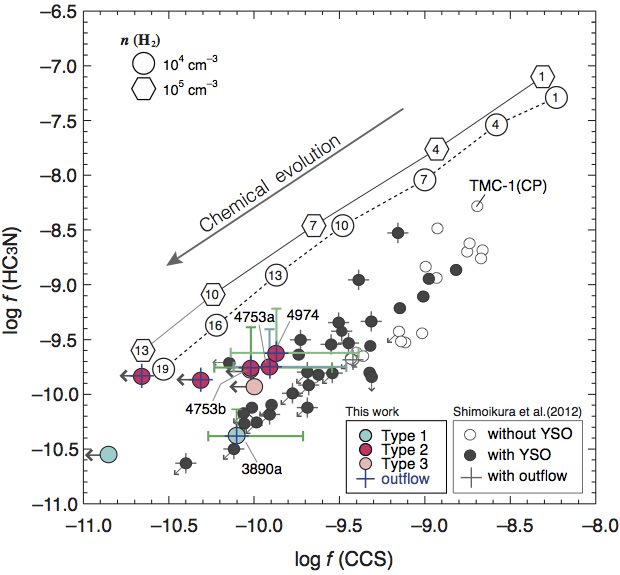}
\caption{
The diagram of ${f}$(HC$_3$N) vs. ${f}$(CCS) of the nine clumps detected in HC$_3$N, 
compared with the sample of cores with and without YSOs reported by \cite{Shimoikura2012}. 
The plots are the geometrical mean of the upper and lower limits of ${f}$(HC$_3$N) and ${f}$(CCS)
arising from the different assumption of $T_{\rm ex}$ (see text).
We show error bars (green line) representing the upper and lower limits only for the clumps detected both in CCS and HC$_3$N.
Arrows indicate the upper limits measured from the noise level.
Larger circles and hexagons represent the fractional abundances calculated by \cite[][see their Figure 14]{Suzuki1992}
for the initial hydrogen densities $n(\rm H_2)=10^{4}$ cm$^{-3}$ and $10^{5}$ cm$^{-3}$, respectively. 
Numbers inside the symbols represent the time in units of 10$^{5}$ yr from the beginning of the chemical reaction. 
\label{fig:CCS}}
\end{center} 
\end{figure*}

\begin{figure*}
\begin{center}
\includegraphics[scale=0.5]{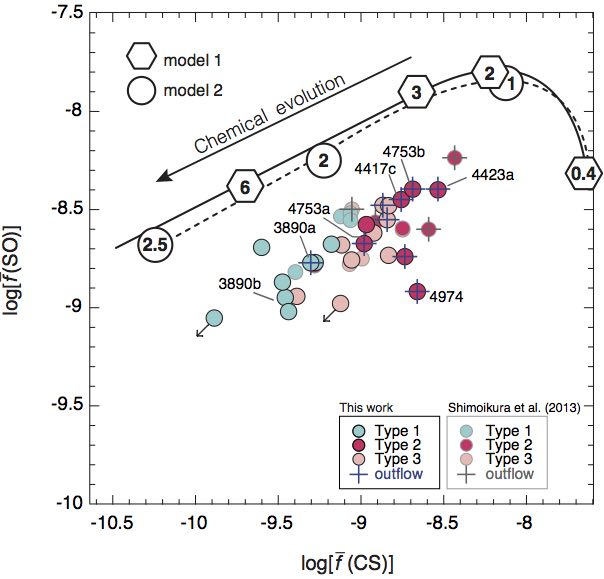}
\caption{
The diagram of $\overline{f}$(SO) and $\overline{f}$(CS) of the 24 clumps.
We also plotted the clumps presented by \cite{Shimoikura2013}. 
Arrows indicate the upper limits.
Larger circles and hexagons represent the fractional abundances calculated for the
two different models called “model 1” and “model 2” by \citet[][see their Figures 2 and 3]{Bergin}. 
Numbers inside the symbols represent the time in units of 10$^{6}$ yr from the beginning of the chemical reaction. 
\label{fig:SO_CS}}
\end{center} 
\end{figure*}

\begin{figure*}
\begin{center}
\includegraphics[scale=.4]{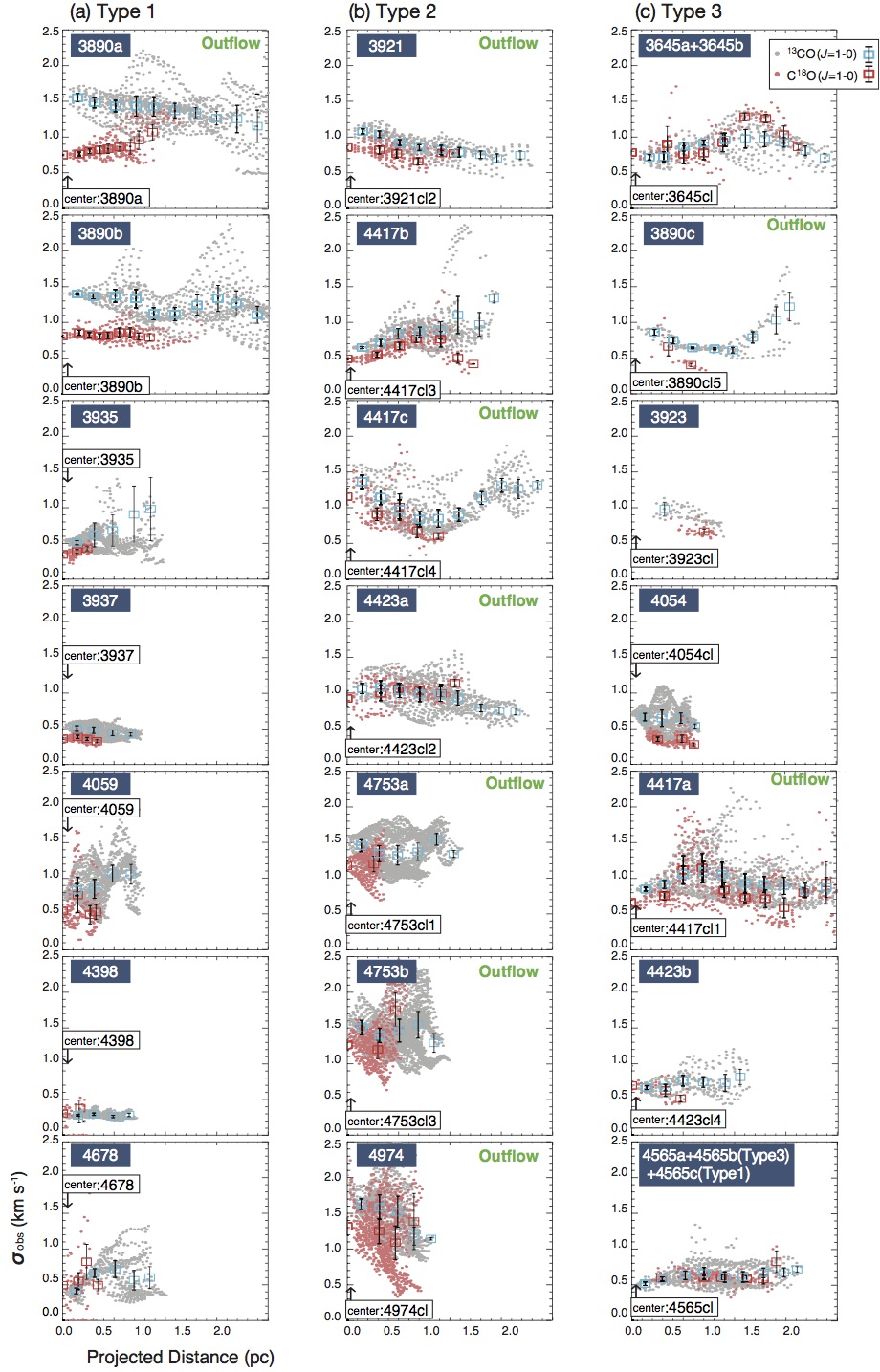}
\caption{
Variation of velocity dispersion (${\sigma_{\rm obs}}$) for each clump as a function of the distance from the cluster center (for the Types 2 and 3 clumps)
or the \eco emission peak position (for the Type1 clumps).
The different color dots represent the velocity dispersion values measured 
from the \tco (grey) and \eco (pink) data. 
Blue and red squares represent the median values of ${\sigma_{\rm obs}}$ in the 0.2 pc bin for the \tco and the \eco data, respectively. 
The error bars show the standard deviation of each data in the bins.
\label{fig:sigma}}
\end{center} 
\end{figure*}

\begin{figure*}
\begin{center}
\includegraphics[scale=0.4]{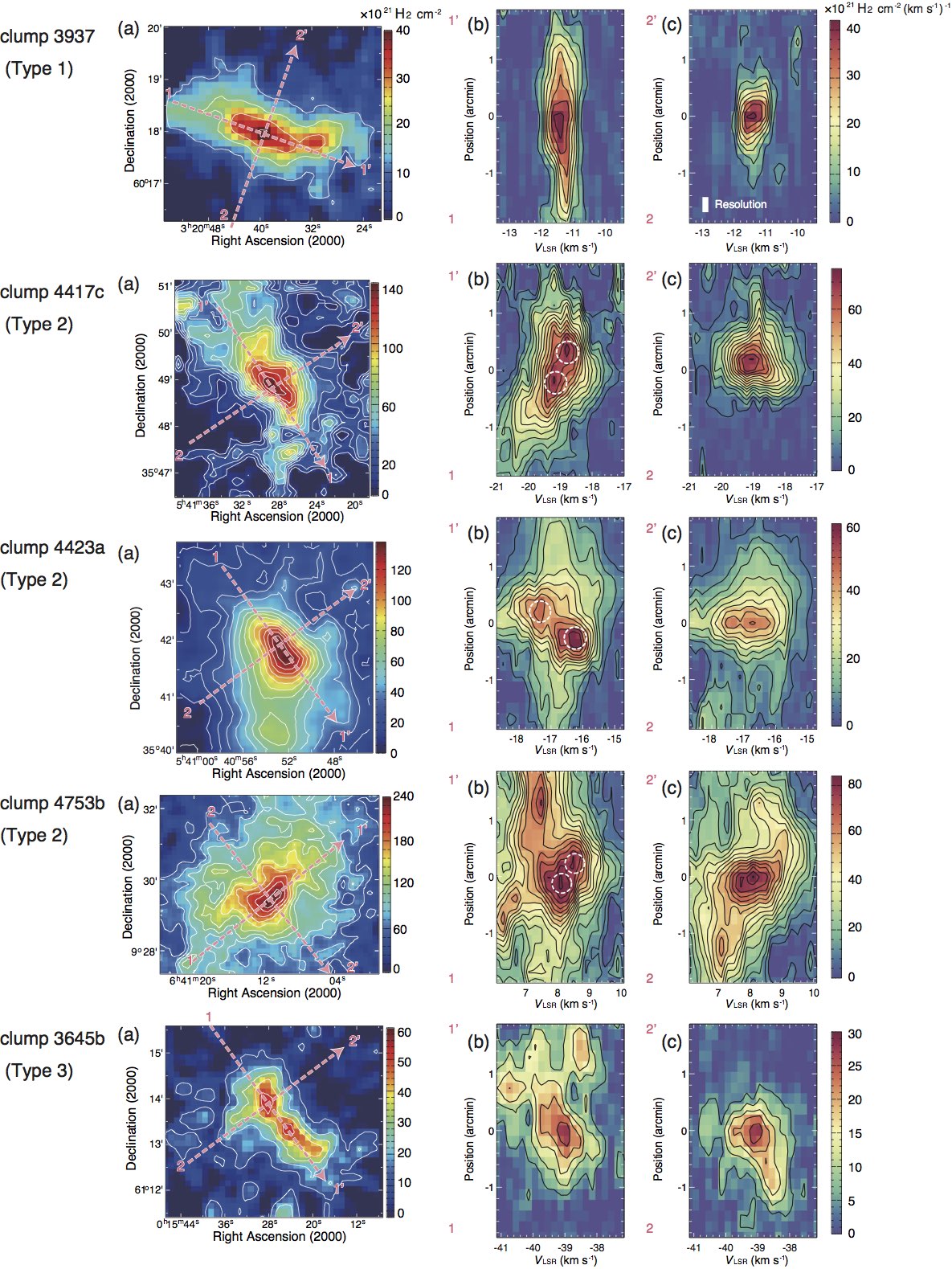}
\caption{
Distributions of the H$_2$ column density derived from the C$^{18}$O integrated intensity,
observed position-velocity diagram measured along the major axis (cut 1-1') , and that measured along the minor axis (cut 2-2'),
are displayed in panels (a)--(c) for clumps classified to Types 1--3.
The lowest contour and contour interval for panel (a) are
$1 \times 10^{22}$ H$_2$ cm$^{-2}$ for all of the clumps except for clump 4753b whose
lowest contour and contour interval are $5 \times 10^{22}$ H$_2$ cm$^{-2}$ and $2 \times 10^{22}$ H$_2$ cm$^{-2}$, respectively.
The lowest contour and contour interval for panels (b) and (c) are $5\times10^{21}$ H$_2$ cm$^{-2}$(km s$^{-1}$)$^{-1}$
for all of the clumps. White broken circles in panels (b) for the Type 2 clumps denote the double-peaked feature (see text).
\label{fig:PV}}
\end{center} 
\end{figure*}

As indicated in Figure \ref{fig:corr1},
these clumps lie in the same loci as Type 2 clumps in the figures,
and they appear to be potential sites of cluster formation in the future. 
The Type 1 clumps are associated with several Class I sources \citep{Jose2016},
and 3890a is associated with a molecular outflow (in Figure \ref{fig:outflow}),
and thus sporadic star formation should be taking place there.
However, the clumps are not associated with any apparent clusters nor they don't show the double-peaked features
in the PV diagrams, suggesting that they are not collapsing but should be gravitationally stable on the clump-scale.
Based on the calculations denoted in Section \ref{sec:corr}, 
we found that the turbulence of the Type 1 clumps is not large enough to support
the clumps by itself (i.e., $M_{\rm vir} < M_{\rm LTE}$), and therefore
additional clump-supporting force is needed. An easy solution for the additional force
is the magnetic field. The clumps would be gravitationally stable if they
are supported by rather strong magnetic field of an order of $\sim1$ mG.
In other words, the clumps may be magnetically sub-critical
\citep{Nakano1985,Shu1987}, i.e., their collapsing motions by the self-gravity are prevented by the magnetic filed.
Observational data compiled by \cite{Crutcher2010} show that the magnetic field of this strength ($\sim1$ mG) is
possible at the density of the clumps ($\sim10^5$ cm$^{-3}$).
We therefore suggest that the massive Type 1 clumps (3890a and 3890b)
have been stable for a long time due to the strong magnetic filed, which may be the reason
why they are chemically older than the Type 2 clumps. 
The clumps have a potential to collapse and evolve into the Type 2 clumps in the future,
whenever the support by the magnetic field is weakened.

Type 2 clumps such as 4423a and 4753b forming clusters must have evolved
from Type 1 clumps which used to be younger in terms of chemical compositions
being located in earlier stages in Figures \ref{fig:CCS} and \ref{fig:SO_CS}
than the observed Type 2 clumps. It is likely that such Type 1 clumps are not found in
our sample. The reason why we don't find them in our sample is probably their short lifetime.
The Type 1 clumps have a typical density of $\sim 1 \times 10^5$ cm$^{-3}$ (Figure \ref{fig:corr1})
corresponding to the fee-fall time of only $\tau_{ff}\simeq10^{5}$ yr, and thus they should contract
as soon as they lose the internal supporting forces such as the turbulence and the magnetic field.
Beside the Type 1 clumps, ages of the clusters forming in the Type 2 clumps are $>1 \times 10^6$ yr,
and therefore the probability to find such Type 1 clumps should be less than 1/10
of that for the Type 2 clumps. This should be the reason why it is difficult to find such Type 1 clumps.

\begin{figure*}
\begin{center}
\includegraphics[scale=.4]{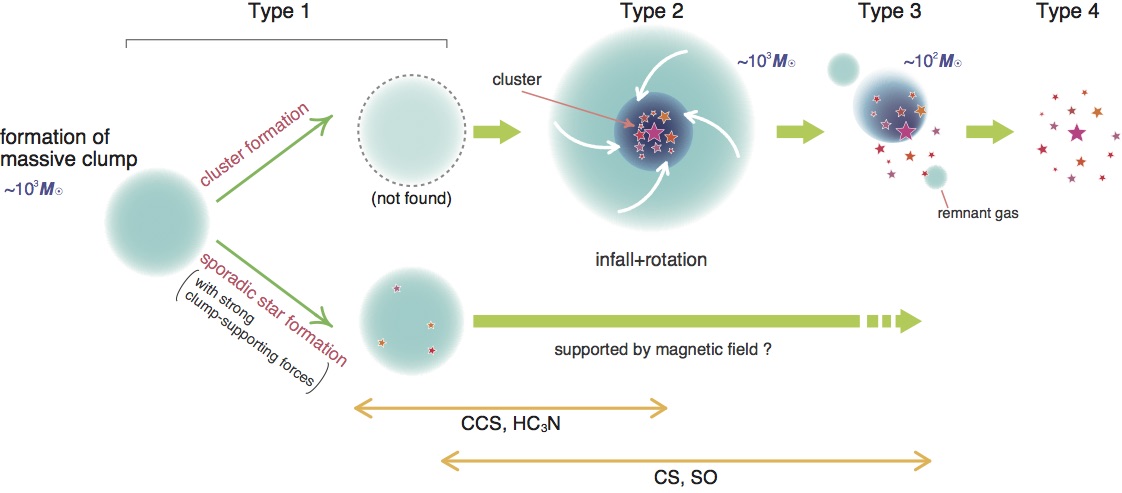}
\caption{
The proposed evolutionary scenario of cluster formation.
\label{fig:evolution}}
\end{center} 
\end{figure*}

All of the above imply how the clumps of different types should evolve:
When massive and dense Type 1 clumps  ($\sim1000$ $M_\sun$, $\sim10^5$ cm$^{-3}$)
are formed, they are initially stable being supported by the turbulence and magnetic field.
They start collapsing and evolve into Type 2 clumps as soon as they lose the clump-supporting forces,
or they remain as Type 1 clumps without collapsing if they maintain the clump supporting forces
(e.g., the magnetic field of an order of $\sim1$ mG).
The Type 2 clumps collapse on the clump-scale, and form clusters at the clump center.
Along with the growth of the clusters, the clumps
disperse by the feedback of cluster formation (e.g., powerful outflows and stellar wind),
which corresponds to our Type 3 category.
At the end, only clusters remain (Type 4).
In Figure \ref{fig:evolution},  we schematically summarize the above evolutionary scenario of the cluster-forming clumps.



\section{CONCLUSIONS}\label{sec:conclusions}

To study cluster formation processes systematically, 
we carried out observations of 15 dust condensations
selected from a catalog compiled by \cite{Dobashi2011}
in various molecular lines 
using the NRO 45 m telescope.
The main results of the observations can be summarized as follows:

\begin{enumerate}

\item
The observations resolved 
the 15 dust condensations into 24 molecular clumps in the \eco data whose masses vary from 5 to $4397 M_\sun$. 
We also analyzed the 2MASS point source catalog to investigate the star density distributions of
the associated clusters, and identified 23 clusters in the observed regions.
Among the 24 clumps,
16 are associated with one ore more IR clusters, 
whereas the other eight are not associated with apparent IR clusters.
We also detected nine outflow sources in the 24 clumps, and  two of the outflows are newly identified by this study.

\item
To assess the relationship between the clumps and IR clusters, 
we investigated the spatial correlations between $N(\rm H_2)$ and the star densities,
and classified the clumps and IR clusters into four types:
eight clumps having no associated IR clusters are classified to Type 1,
seven clumps showing good correlations with IR clusters are classified to Type 2,
nine clumps showing poor correlations with IR clusters are classified to Type 3,
and
six IR clusters with no associated clumps are classified to Type 4.


\item
The Type 2 clumps have a typical mass of $\sim1 \times 10^{3} M_\sun$,
a radius of $\sim0.5$ pc, and a density of $\sim1 \times 10^5$ cm$^{-3}$,
and the distributions of the associated clusters coincide well with the \eco distributions, 
indicating that the clusters are forming around the center of the clumps. 
The velocity dispersion in the Type 2 clumps tends to be larger ($>0.5$ \kms) at the cluster center. 
We found that such Type 2 clumps are accompanied by two velocity components 
that differ by $\sim1$ \kms, which can be recognized as the two peaks in the position-velocity diagram. 
On the other hand, neither Type 1 nor Type 3 clumps exhibit such velocity structures. 
In our previous paper, 
we presented an evidence that the clump 4423a (Type 2) showing the two velocity components is infalling with rotation
on the clump-scale.
In this paper, we found at least two similar cases (4417c and 4753b) in the other Type 2 clumps,
which indicates that the infalling motion with rotation is a common phenomenon among the Type 2 clumps.

\item
We suggest that the clump + IR cluster systems should evolve from Type 1 to Type 4.
The Type 1 clumps are stable in the beginning and evolve into Type 2 clumps as soon as they lose the clump-supporting forces,
and the Type 2 clumps are collapsing on the clump-scale to form clusters at the clump center. 
The Type 3 clumps are in a more advanced stage with much less molecular gas,
and the Type 4 clusters are more evolved and their natal clumps have already dispersed.
All of the clumps classified to Type 1 in this study are older than the Type 2 clumps in terms of chemical compositions.
We suggest that they are the Type 1 clumps being gravitationally stable without collapsing for a long time
due to the strong magnetic field, or that they are the remnants of Type 3 clumps.
Type 1 clumps which are younger than the observed Type 2 clumps in terms of chemical composition
should be rare to find due to their short lifetime.

\end{enumerate}



\acknowledgments
This work was financially supported by Grant-in-Aid for Scientific Research 
(Nos. 17H02863, 17H01118, 26287030, 17K00963)
of Japan Society for the Promotion of Science (JSPS). 
The 45 m radio telescope is operated by NRO, a branch of NAOJ. 
\clearpage








\floattable
\begin{deluxetable*}{lccc} 
\tablecaption{List of Observed Dust Condensations \label{tab:chara}} 
\tablehead{ 
\colhead{Dust Condensation No.\tablenotemark{(1)}}	& 	\colhead{IR cluster\tablenotemark{a}}& 	
 \colhead{Associated Objects} 	& 	\colhead{Distance (kpc)}  
}
\startdata 
3645	&	Y		&	S172				&	2.4\tablenotemark{(2)} 	\\
3890	&	Y		&	S190, AFGL333, IC1805-W, BRC5	&	2.0\tablenotemark{(3)} 	\\
3921	&	Y		&	S199, AFGL4029			&	2.1\tablenotemark{(4)} 	\\
3923	&	Y		&	S201						&	2.1\tablenotemark{(4)} 	\\
3935	&	\nodata	&	\nodata				&	0.9 (For S202\tablenotemark{(2)})	\\
3937	&	\nodata	&	\nodata				&	0.9 (For S202\tablenotemark{(2)} )	\\
4054	&	Y		&	BFS32				&	0.6\tablenotemark{(2)} 	\\
4059	&	\nodata	&	\nodata				&	0.6 (For BFS52\tablenotemark{(2)} )	\\
4398	&	\nodata	&	\nodata				&	0.81\tablenotemark{b} 	\\
4417	&	Y		&	S235					&	1.56\tablenotemark{(5)} 	\\
4423	&	Y		&	S235A, B				&	1.56\tablenotemark{(5)} 	\\
4565	&	Y		&	BFS51				&	1.6\tablenotemark{(6)} 	\\
4678	&	\nodata	&	\nodata				&	1.0(For S268\tablenotemark{(2)} )	\\
4753	&	Y		&	Mon OB1-D, NGC2264C, D	&	0.738\tablenotemark{(7)} 	\\
4974	&	Y		&	Mon R2			&	0.893\tablenotemark{(8)} 	\\
\enddata 

\tablenotemark{a} {See Table \ref{tab:clump+cluster} for the references of the IR clusters.}

\tablenotemark{b} {For No.4398, we calculated the kinematic distance using the $^{13}$CO peak velocity obtained with the 45 m observations ($V_{\rm {LSR}}=-2.2$ km s$^{-1}$). 
The distance is calculated using the Galactic rotation model of \cite{Kerr},\cite{Wouterloot1989}, and \cite{Wouterloot1990}, 
assuming the distance of the Sun from Galactic Center $R_{0} = 8.5$ kpc
and the rotation speed of the Galaxy $\Theta_{0}$ = 220 km s$^{-1}$ at $R_{0}$.}

\tablerefs{
\tablenotemark{(1)} \cite{Dobashi2011}
\tablenotemark{(2)} \cite{Russeil}
\tablenotemark{(3)} \cite{Xu2006}
\tablenotemark{(4)} \cite{Chauhan}
\tablenotemark{(5)} \cite{Burns2015}
\tablenotemark{(6)} \cite{carpenter1990}
\tablenotemark{(7)} \cite{Kamezaki2014}
\tablenotemark{(8)} \cite{Dzib2016}
}

\end{deluxetable*}


\begin{deluxetable*}{lcrcc} 
\tablecaption{NRO 45m Observations \label{tab:45obs}} 
\tablehead{ 
 \colhead{Molecule} & \colhead{Transition}   & \colhead{Frequency}  & \colhead{Beam} &  \colhead{$\Delta T_{\rm mb}$} \\
 \colhead{} &\colhead{}  & \colhead{(GHz)}   & \colhead{(arcsec)} & \colhead{(K)} 
}
\startdata 
CCS			&	$J_{N}$=4$_{3}-3_{2}$	&	45.379033	&	$36\farcs8$	&	$0.1-0.3$	\\
HC$_{3}$N	&	$J=5-4$				&	45.490316	&	$36\farcs7$	&	$0.1-0.3$	\\
C$^{34}$S	&	$J=2-1$				&	96.412950	&	$17\farcs3$	&	$0.3-0.6$	\\
CS			&	$J=2-1$				&	97.980953 	&	$17\farcs0$	&	$0.4-0.7$	\\
SO			&	$J_{N}$=2$_{3}$--1$_{2}$	&	99.299905 	&	$16\farcs8$	&	$0.3-0.6$	\\
C$^{18}$O	&	$J=1-0$				&	109.782176 	&	$15\farcs2$	&	$0.4-0.8$	\\
$^{13}$CO	&	$J=1-0$				&	110.201354 	&	$15\farcs1$	&	$0.5-0.9$	\\
$^{12}$CO	&	$J=1-0$				&	115.271202 	&	$14\farcs5$	&	$1.2-1.8$	\\
\enddata 
\tablecomments{The rest frequency for the CCS line is taken from \cite{Yamamoto1990}, and those of the other lines are taken
from the website of \cite{Lovas}.}

\end{deluxetable*} 


\floattable
\begin{deluxetable*}{lrrcccrrrrrrrr} 
\tabletypesize{\scriptsize}
\setlength{\tabcolsep}{0.02in} 
\tablewidth{0pc} 
\tablecaption{Physical Properties of the Clumps \label{tab:mass}} 
\tablehead{ 
 \colhead{} &	 \colhead{$\alpha$(J2000)\tablenotemark{a}} 	&	\colhead{$\delta$(J2000)\tablenotemark{a}}	&	\colhead{$R$} 	&	\colhead{$\overline{\Delta V}$(C$^{18}$O)\tablenotemark{b}}	&	\colhead{$\overline{T_{\rm ex}}$\tablenotemark{b}}	&	\colhead{$M_{\rm LTE}$}	&	\colhead{$M_{\rm vir}$}	&	 \colhead{$n({\rm H_2})$\tablenotemark{b}}	&	\colhead{$\overline{N}({\rm H_2})$\tablenotemark{b}}	&	\colhead{$\overline{N}$(C$^{18}$O)\tablenotemark{b}}	&	\colhead{$\overline{f}$(SO)\tablenotemark{b}} 	&	\colhead{$\overline{f}$(CS)\tablenotemark{b}} 	&	\colhead{$\overline{f}$(C$^{34}$S)\tablenotemark{b}} 	\\
 \colhead{Clump} 	&	\colhead{($^h\:^m\:^s$)}	&	\colhead{(${\arcdeg}\:{\arcmin}\:\:{\arcsec}$)} 	&	\colhead{(pc)}	&	\colhead{(km s$^{-1}$)} 	&	\colhead{ (K)}	&	\colhead{($M_{\sun}$)}	&	\colhead{($M_{\sun}$)}	&	\colhead{($10^{4}$cm$^{-3}$)} 	&	\colhead{(10$^{22}$cm$^{-2}$)}	&	\colhead{(10$^{15}$cm$^{-2}$)} 	&	\colhead{(10$^{-9}$)} 	&	\colhead{(10$^{-9}$)} 	&	\colhead{(10$^{-10}$)} 	
}
\startdata
3645a	&	00:15:25.0	&	61:13:17		&	0.43 	&	2.7 	&	35.8 	&	398  	&	681 	&	1.7  	&	3.9 	&	6.3 	&	1.7 	&	0.9 	&	$<$1.1	\\
3645b	&	00:15:29.2	&	61:14:02		&	0.41 	&	1.9 	&	38.4 	&	474  	&	347 	&	2.4  	&	3.6 	&	5.8 	&	1.8 	&	1.5 	&	1.7 	\\
3890a	&	02:28:03.1	&	61:27:39		&	0.61 	&	2.1 	&	36.0 	&	3432  &	573 	&	5.2  	&	13.5 	&	23.6 	&	1.7 	&	0.5 	&	2.8 	\\
3890b	&	02:28:04.1	&	61:29:16		&	0.62 	&	2.4 	&	35.6 	&	4397 &	721 	&	6.4  	&	17.3 	&	30.4 	&	1.1 	&	0.3 	&	2.2 	\\
3890c	&	02:28:57.7	&	  61:33:01		&	0.23 	&	1.1 	&	35.9 	&	113  	&	56 	&	3.2  	&	3.0 	&	4.7 	&	3.3 	&	1.4 	&	$<$2.0	\\
3921		&	03:01:30.4	&	60:29:11		&	0.57 	&	1.9 	&	50.5 	&	1508  &	439 	&	2.8  	&	6.7 	&	11.4 	&	1.8 	&	1.8 	&	2.0 	\\
3923		&	03:03:22.5	&	  60:28:03		&	0.28 	&	1.6 	&	43.0 	&	249  	&	141 	&	3.9  	&	4.6 	&	7.7 	&	2.4 	&	1.2 	&	$<$0.6	\\
3935		&	03:15:41.4	&	 60:02:23		&	0.19 	&	0.9 	&	28.8 	&	25  	&	35 	&	1.2  	&	2.7 	&	4.2 	&	1.3 	&	0.3 	&	$<$0.3	\\
3937		&	03:20:38.9	&	  60:17:54		&	0.21 	&	0.9 	&	26.4 	&	86  	&	32 	&	3.2  	&	2.8 	&	4.4 	&	1.0 	&	0.4 	&	$<$0.6	\\
4054		&	03:51:42.7	&	51:27:52		&	0.16 	&	1.0 	&	34.9 	&	48  	&	31 	&	4.0 	&	2.7 	&	4.1 	&	1.1 	&	0.4 	&	$<$0.4	\\
4059		&	03:56:50.4	&	  51:47:19		&	0.20 	&	1.3 	&	24.5 	&	71  	&	67 	&	3.1  	&	2.5 	&	3.9 	&	1.7 	&	0.5 	&	$<$0.8	\\
4398		&	05:14:09.6	&	32:45:07		&	0.10 	&	0.9 	&	34.2 	&	5  	&	14 	&	1.6 	&	1.0 	&	1.2 	&	$<$0.9	&	$<$0.1	&	$<$0.9	\\
4417a	&	05:40:59.7	&	  35:49:20		&	0.64 	&	3.1 	&	54.4 	&	1852  &	1464 &	2.4  	&	6.5 	&	11.1 	&	2.8 	&	1.4 	&	1.3 	\\
4417b	&	05:41:26.9	&	  35:52:12		&	0.46 	&	1.5 	&	54.8 	&	994  	&	253 	&	3.5  	&	7.6 	&	13.1 	&	2.7 	&	1.1 	&	$<$0.7	\\
4417c	&	05:41:29.3	&	  35:48:57		&	0.40 	&	2.5 	&	56.5 	&	993  &	609 	&	5.4  	&	8.7 	&	15.0 	&	3.6 	&	1.7 	&	1.7 	\\
4423a	&	05:40:52.9	&	  35:41:35		&	0.41 	&	2.5 	&	47.7 	&	955  	&	617 	&	4.8  	&	7.9 	&	13.5 	&	4.0 	&	2.9 	&	4.3 	\\
4423b	&	05:40:53.5	&	  35:38:28		&	0.25 	&	1.5 	&	55.7 	&	177  	&	142 	&	3.9  	&	4.2 	&	6.8 	&	$<$1.1	&	$<$0.8	&	$<$1.2	\\
4565a	&	06:18:36.7	&	  23:18:32		&	0.19 	&	1.7 	&	58.1 	&	131  	&	115 	&	6.6  	&	5.3 	&	8.8 	&	3.3 	&	1.5 	&	$<$1.2	\\
4565b	&	06:18:42.7	&	23:20:02		&	0.31 	&	1.6 	&	38.9 	&	331  	&	172 	&	3.8  	&	5.1 	&	8.5 	&	2.1 	&	0.8 	&	$<$1.0	\\
4565c	&	06:18:46.0	&	23:21:17		&	0.22 	&	1.4 	&	32.9 	&	161  	&	92 	&	5.2  	&	3.7 	&	6.0 	&	2.1 	&	0.7 	&	$<$0.8	\\
4678		&	06:09:52.5	&	  13:44:48		&	0.14 	&	1.4 	&	36.7 	&	26  	&	59 	&	3.3  	&	2.1 	&	3.1 	&	2.0 	&	0.3 	&	$<$0.9	\\
4753a	&	06:41:05.0	&	   09:35:15	&	0.27 	&	2.9 	&	50.6 	&	1123  &	486 	&	19.8  	&	20.6 	&	36.5 	&	2.1 	&	1.0 	&	1.2 	\\
4753b	&	06:41:10.6	&	   09:29:38	&	0.35 	&	3.3 	&	50.8 	&	1203  &	810 	&	9.7 	&	13.6 	&	23.8 	&	4.0 	&	2.0 	&	2.1 	\\
4974		&	06:07:44.4	&	  $-$06:23:26	&	0.50 	&	3.3 	&	59.4 	&	2581  &	1076&	7.2  	&	13.3 	&	23.3 	&	1.2 	&	2.2 	&	1.7 	\\
\enddata

\tablenotemark{a} {Peak position of the C$^{18}$O ($J=1-0$) integrated intensity in the equatorial coordinates.}

\tablenotemark{b} {The average value of the clumps.}
 
\tablecomments{For the fractional abundance of SO, we calculated the mean partition function $Q$ to be $76.3-176.9$ for the assumed $T{\rm_{ex}}$ of $24.5 - 82.1$ K using the parameters measured by \cite{Tiemann}.
}
\end{deluxetable*} 

\clearpage


\begin{deluxetable*}{ccccccccc} 
\tablecaption{Physical Properties Derived from CCS and ${\rm HC_{3}N}$ \label{tab:ccs}} 
\tablehead{ 
 \colhead{} 	&\colhead{$\alpha$(J2000)} 	&	\colhead{$\delta$(J2000)}	&
 \colhead{$T_{\rm ex}$\tablenotemark{a}}	& \colhead{$N({\rm H_2})$} & 
 \colhead{$N({\rm CCS})$\tablenotemark{b}}	&\colhead{$N({\rm HC_{3}N})$\tablenotemark{b}}	 &
 \colhead{$f$(CCS)\tablenotemark{b}}  &\colhead{$f({\rm HC_{3}N})$\tablenotemark{b}}\\
  \colhead{Clump} 	&	\colhead{($^h\:^m\:^s$)}	&	\colhead{(${\arcdeg}\:{\arcmin}\:\:{\arcsec}$)} 	&	\colhead{(K)}	&	\colhead{(10$^{22}$cm$^{-2}$)}	&	\colhead{(10$^{12}$cm$^{-2}$)}	&	\colhead{(10$^{12}$cm$^{-2}$)}	&	\colhead{(10$^{-10}$)} 	&	\colhead{(10$^{-10}$)} 		
 }
\startdata
3645a	&	00:15:30.1	&	61:14:30	&	35.9 	&	2.1 	&	$<1.4-<4.9$	&	$4.7-6.5$	&	$<0.7-<2.3$	&	$2.2-3.1$	\\
3890a	&	02:28:4.9		&	61:27:45	&	36.2 	&	15.6 	&	$8.6-29.7$	&	$8.5-11.7$	&	$0.5-1.9$	&	$0.5-0.7$	\\
3890b	&	02:28:7.0		&	61:29:15	&	34.2 	&	19.1 	&	$<3.0-<6.4$	&	$6.9-9.5$	&	$<0.2-0.3$	&	$0.4-0.5$	\\
3921		&	03:01:30.0	&	60:29:15	&	52.1 	&	6.9 	&	$<1.7-<9.4$	&	$11.5-21.2$	&	$<0.2-1.4$	&	$1.7-3.1$	\\
4423a	&	05:40:53.0	&	35:41:35	&	47.8 	&	8.9 	&	$<4.5-<5.4$	&	$16.1-27.9$	&	$<0.5-0.6$	&	$1.8-3.1$	\\
4565b	&	06:18:43.3	&	23:19:45	&	37.4 	&	4.0 	&	$<2.6-<9.2$	&	$6.0-8.5$	&	$<0.7-<2.3$	&	$1.5-2.1$	\\
4753a	&	06:40:58.4	&	09:36:8	&	50.2 	&	13.4 	&	$10.1-46.5$	&	$30.2-53.9$	&	$0.8-3.5$	&	$2.2-4.0$	\\
4753b	&	06:41:10.6	&	09:29:23	&	55.3 	&	18.2 	&	$10.3-50.9$	&	$38.8-75.1$	&	$0.6-2.8$	&	$2.1-4.1$	\\
4974		&	06:07:39.0	&	-06:22:60	&	66.1 	&	14.2 	&	$10.1-57.6$	&	$38.6-87.4$	&	$0.7-4.1$	&	$2.7-6.2$	\\
\enddata
\tablecomments{The parameters were derived at the peak position of the HC$_3$N integrated intensity.
To estimate the column densities of CCS, 
we calculated $Q$ to be $24.302-651.75970$ for the assumed $T{\rm_{ex}}$ of $5.0 - 66.1$ K using the parameters measured by \cite{Yamamoto1990}.
}
\tablenotemark{a} {$T_{\rm ex}$ calculated using the $^{12}$CO data.}

\tablenotemark{b} {Minimum and maximum estimates for parameters assuming a flat  $T_{\rm ex}$ of 5 K and
that calculated using the $^{12}$CO data, respectively.}

\end{deluxetable*} 

\clearpage


\begin{deluxetable*}{cccc} 
\tabletypesize{\scriptsize}
\tablecaption{Derived Velocity Dispersion \label{tab:disper}} 
\tablehead{ 
	\colhead{Clump} & \colhead{${\sigma_{\rm obs(C^{18}O)}}$}  &\colhead{${\sigma_{\rm therm}}$} &\colhead{${\sigma_{\rm 3D}}$}\\
	\colhead{} 	& \colhead{(km s$^{-1}$)} 	&	\colhead{(km s$^{-1}$)} 	&	\colhead{(km s$^{-1}$)} 	
}
\startdata
3645a	&	1.16 	&	0.10 	&	1.96 	\\
3645b	&	0.77 	&	0.10 	&	1.32 	\\
3890a	&	0.91 	&	0.10 	&	1.55 	\\
3890b	&	0.82 	&	0.10 	&	1.42 	\\
3890c	&	0.48 	&	0.10 	&	0.84 	\\
3921	&	0.78 	&	0.12 	&	1.33 	\\
3923	&	0.66 	&	0.11 	&	1.14 	\\
3935	&	0.36 	&	0.09 	&	0.60 	\\
3937	&	0.36 	&	0.09 	&	0.60 	\\
4054	&	0.31 	&	0.09 	&	0.51 	\\
4059	&	0.59 	&	0.08 	&	1.00 	\\
4398	&	0.27 	&	0.10 	&	0.44 	\\
4417a	&	0.87 	&	0.12 	&	1.49 	\\
4417b	&	0.70 	&	0.12 	&	1.10 	\\
4417c	&	0.86 	&	0.12 	&	1.53 	\\
4423a	&	1.02 	&	0.11 	&	1.75 	\\
4423b	&	0.62 	&	0.12 	&	1.06 	\\
4565a	&	0.63 	&	0.13 	&	1.07 	\\
4565b	&	0.61 	&	0.10 	&	1.04 	\\
4565c	&	0.59 	&	0.10 	&	1.00 	\\
4678	&	0.45 	&	0.10 	&	0.81 	\\
4753a	&	1.19 	&	0.12 	&	2.05 	\\
4753b	&	1.35 	&	0.12 	&	2.33 	\\
4974	&	1.21 	&	0.13 	&	2.07 	\\
\enddata

\end{deluxetable*}

\floattable
\begin{deluxetable*}{cccccccccccccc} 
\tabletypesize{\scriptsize}
\setlength{\tabcolsep}{0.01in} 
\tablecaption{Derived Parameters of the Outflows. \label{tab:outflow}} 
\tablehead{ 
\colhead{Clump}	&	\colhead{wing}	&	 \colhead{$\overline{T_{\rm ex}}$} 	&	\colhead{$V_{\rm sys}$}	&	\colhead{$V_{\rm char}$}	
&	\colhead{$V_{\rm range}$}	&	\colhead{$S_{\rm lobe}$}	&	\colhead{$R_{\rm lobe}$}	&	\colhead{$\int \int T_{\rm mb}\,{\rm d}V {\rm d}S$}	
&	\colhead{$t_{\rm d}$}	&	\colhead{$\tau_{\rm max}$}	&	\colhead{$M_{\rm lobe}$\tablenotemark{a}}	&	\colhead{$P_{\rm lobe}$ \tablenotemark{a}}			
&	\colhead{$\dot{P}_{\rm lobe}$\tablenotemark{a}}\\
\colhead{}	&	\colhead{}	&	\colhead{(K)}	&	\colhead{(kms$^{-1}$)}	&	\colhead{(kms$^{-1}$)}	&	\colhead{(kms$^{-1}$)}	
&	\colhead{(arcsec$^{2}$)}	&	\colhead{(pc)}	&	 \colhead{(K km s$^{-1}$arcmin$^2$)} 	&	\colhead{(10$^{4}$yr)}	
&	\colhead{}	&	\colhead{($M_{\sun}$)}&	 \colhead{($M_{\sun}$ km s$^{-1}$)}					
&	\colhead{(10$^{-4}$$M_{\sun}$ km s$^{-1}$yr$^{-1}$)}			
}
\startdata
3890a & blue & 35.5 & $-$48.5 & 9.2 & ($-$59,$-$53.5) & 2981 & 0.30 & 8.8 & 3.2 & 1.7 & 1.3 - 2.7 & 13.3 - 27.4 & 4.5 - 9.4 \\
 & red & 35.3 &  & 8.0 & ($-$44,$-$40) & 2756 & 0.29 & 7.0 & 3.5 & 2.4 & 1.1 - 2.8 & 8.8 - 23.3 & 2.6 - 6.9 \\
3890c & blue & 29.3 & $-$51.5 & 5.4 & ($-$59, $-$53) & 4838 & 0.38 & 17.8 & 7.0 & 1.5 & 2.5 - 4.8 & 14.8 - 28.8 & 2.4 - 4.6 \\
 & red & 30.0 &  & 6.2 & ($-$49, $-$44) & 3544 & 0.33 & 9.2 & 5.1 & 6.7 & 1.3 - 8.6 & 7.3 - 48.9 & 1.3 - 8.7 \\
3921 & blue & 47.6 & $-$38.1 & 7.3 & ($-$46, $-$41) & 10744 & 0.60 & 33.7 & 8.0 & 6.4 & 7.7 - 49.6 & 69.9 - 450.2 & 10.8 - 69.9 \\
 & red & 44.3 &  & 6.4 & ($-$36, $-$30) & 33019 & 1.05 & 166.6 & 16.0 & 9.9 & 36.4 - 361.1 & 271.0 - 2685.5 & 19.7 - 195.0 \\
4417a & blue & 58.1 & $-$20.3 & 7.7 & ($-$28, $-$24) & 5738 & 0.32 & 20.2 & 4.1 & 3.4 & 3.0 - 10.4 & 22.5 - 79.2 & 5.4 - 19.0 \\
4417c & blue & 55.6 &  & 7.0 & ($-$28, $-$24) & 2925 & 0.23 & 8.9 & 3.2 & 5.6 & 1.2 - 6.8 & 7.9 - 44.3 & 2.3 - 12.7 \\
4423a & blue & 53.2 & $-$16.9 & 7.7 & ($-$24, $-$20) & 3319 & 0.24 & 15.5 & 3.1 & 2.5 & 2.1 - 5.7 & 17.3 - 47.8 & 6.0 - 16.6 \\
4753a & blue & 44.9 & 5.3 & 11.3 & ($-$9, 1) & 13219 & 0.23 & 81.9 & 2.0 & 2.8 & 2.1 - 6.3 & 25.0 - 75.5 & 13.2 - 39.8 \\
4753b & blue & 53.6 & 7.9 & 12.0 & ($-$6, 2) & 787 & $<$0.1 & 3.2 & $<$0.5 & 3.7 & $<$0.1 - 0.2 & 0.4 - 1.6 & $<$0.1 - 3.1 \\
 & red & 52.8 &  & 11.3 & (13, 20) & 7031 & 0.17 & 53.3 & 1.5 & 3.8 & 1.5 - 6.0 & 16.5 - 63.8 & 10.5 - 40.8 \\
4974 & blue & 65.0 & 10.4 & 10.7 & (0, 5.5) & 24413 & 0.38 & 156.6 & 3.5 & 8.1 & 8.1 - 65.5 & 82.4 - 670.4 & 22.5 - 183.1 \\
 & red & 55.4 &  & 11.6 & (16.5, 22) & 40894 & 0.50 & 377.8 & 4.2 & 8.4 & 16.9 - 141.7 & 198.1 - 1665.4 & 48.0 - 403.8 \\
\enddata

\tablecomments{\tablenotemark{a} 
Minimum and maximum estimates for parameters of the outflows, which are estimated using $\tau=0$ and $\tau=\tau_{\rm max}$, respectively (see Appendix B).}
\end{deluxetable*} 

\clearpage

\begin{deluxetable*}{crrrrcc} 
\tabletypesize{\scriptsize}
\tablewidth{0pc} 
\tablecaption{Physical Properties of the IR Clusters \label{tab:cluster}} 
\tablehead{ 
 \colhead{IR Cluster} 	&	\colhead{$\alpha$(J2000)}	&	\colhead{$\delta$(J2000)}	&	 \colhead{$R_\star$} 	&	 \colhead{$N_\star$} 			&
 \colhead{Peak density} 	  &	 Dmin/Dmaj \tablenotemark{a}	\\
 \colhead{}	&	\colhead{($^h\:^m\:^s$)}	&	\colhead{(${\arcdeg}\:{\arcmin}\:\:{\arcsec}$)}	&	(pc)	&	\colhead{}		&	{{(stars arcmin$^{-2}$)}}	&	\colhead{}	
}
\startdata
3645cl	&	00:15:29.2	&	61:14:25	&	0.64 	&	63 	&	34 	&	0.5 	\\
3890cl1	&	02:28:16.1	&	61:31:10	&	0.76 	&	172 	&	47 	&	0.4 	\\
3890cl2	&	02:28:20.7	&	61:28:27	&	0.35 	&	48  	&	63 	&	0.8 	\\
3890cl3	&	02:28:30.4	&	61:27:17	&	0.30 	&	29 	&	39 	&	0.9 	\\
3890cl4	&	02:28:47.7	&	61:33:27	&	0.35 	&	43 	&	43 	&	0.5 	\\
3890cl5	&	02:29:04.1	&	61:33:24	&	0.44 	&	69 	&	59 	&	0.7 	\\
3921cl1	&	03:01:19.3	&	60:29:28	&	0.49 	&	59 	&	35 	&	0.8 	\\
3921cl2	&	03:01:33.8	&	60:29:13	&	0.45 	&	57 	&	49 	&	0.5 	\\
3923cl	&	03:03:15.3	&	60:27:58	&	0.40 	&	63 	&	44 	&	0.8 	\\
4054cl	&	03:51:33.3	&	51:29:59	&	0.13 	&	18 	&	24 	&	0.6 	\\
4417cl1	&	05:41:06.9	&	35:49:24	&	0.28 	&	33 	&	33 	&	0.5 	\\
4417cl2	&	05:41:10.0	&	35:50:09	&	0.27 	&	23 	&	23 	&	0.5 	\\
4417cl3	&	05:41:23.9	&	35:52:09	&	0.34 	&	27 	&	25 	&	0.6 	\\
4417cl4	&	05:41:30.8	&	35:49:53	&	0.40 	&	47 	&	23 	&	0.4 	\\
4423cl1	&	05:40:54.9	&	35:44:14	&	0.35 	&	40 	&	27 	&	0.8 	\\
4423cl2	&	05:40:53.1	&	35:42:14	&	0.40 	&	135 	&	65 	&	0.6 	\\
4423cl3	&	05:40:55.5	&	35:40:07	&	0.43 	&	80 	&	33 	&	0.6 	\\
4423cl4	&	05:40:52.4	&	35:38:22	&	0.25 	&	12 	&	21 	&	0.9 	\\
4565cl	&	06:18:38.1	&	23:19:18	&	0.47 	&	94 	&	35 	&	0.6 	\\
4753cl1	&	06:41:04.2	&	09:35:45	&	0.19 	&	36 	&	22 	&	0.5 	\\
4753cl2	&	06:41:06.7	&	09:34:07	&	0.12 	&	15 	&	16 	&	0.7 	\\
4753cl3	&	06:41:09.7	&	09:29:37	&	0.13 	&	34 	&	40 	&	0.6 	\\
4974cl	&	06:07:46.0	&	$-06:22:55$	&	0.28 	&	380 &	86 	&	0.7 	\\
\enddata

\tablenotemark{a} Ratio of the minor axis to the major axis of the cluster surface area fitted with the 2D Gaussian function.
\end{deluxetable*} 

\clearpage

\begin{deluxetable*}{ccccccc} 
\tabletypesize{\scriptsize}
\tablewidth{0pc} 
\tablecaption{Clumps and IR Clusters \label{tab:clump+cluster}} 
\tablehead{ 
 \colhead{Dust Condensation No.} 	&	\colhead{Clump}	&	\colhead{IR Cluster\tablenotemark{a}}	&
 \colhead{Type}	&	\colhead{Outflow\tablenotemark{a}}	&	\colhead{\it{IRAS} \tablenotemark{b}}	&	\colhead{$L(L_\sun$) \tablenotemark{c}}
}
\startdata
3645	&	3645a	&	3645cl	\tablenotemark{(1)}	&	3	&	N	&	\nodata	&	\nodata\\
	&	3645b	&	3645cl	\tablenotemark{(1)}	&	3	&	N	&	00127+6058		&	$4.0\times10^3$\\
3890	&	3890a	&	N		&	1	&	Y\tablenotemark{(5)}	&	\nodata	&	\nodata\\
	&	3890b	&	N		&	1	&	N	&	\nodata	&	\nodata\\
	&	\nodata	&	3890cl1	\tablenotemark{(1)}	&	4	&	N	&	\nodata	&	\nodata	\\
	&	\nodata	&	3890cl2	\tablenotemark{(2)}	&	4	&	N	&	02245+6115		&	$1.0\times10^4$	\\
	&	\nodata	&	3890cl3	\tablenotemark{$\dagger$}	&	4	&	N	&	\nodata	&	\nodata\\
	&	\nodata	&	3890cl4	\tablenotemark{$\dagger$}	&	4	&	N	&	\nodata	&	\nodata	\\
	&	3890c	&	3890cl5	\tablenotemark{(1)}	&	3	&	Y\tablenotemark{(6)}	&	\nodata	&	\nodata\\
3921	&	\nodata	&	3921cl1	\tablenotemark{(2)}	&	4	&	N	&	\nodata	&	\nodata	\\
	&	3921		&	3921cl2	\tablenotemark{(2)}	&	2	&	Y\tablenotemark{(6)}	&	02575+6017	&	$1.3\times10^4$\\
3923	&	3923		&	3923cl	\tablenotemark{(2)}	&	3	&	N	&	02593+6016	&	$1.7\times10^4$\\
3935	&	3935		&	N		&	1	&	N	&	\nodata	&	\nodata\\
3937	&	3937		&	N		&	1	&	N	&	\nodata	&	\nodata\\
4054	&	4054		&	4054cl	\tablenotemark{(1)}	&	3	&	N	&	\nodata	&	\nodata\\
4059	&	4059		&	N		&	1	&	N	&	\nodata	&	\nodata\\
4398	&	4398		&	N		&	1	&	N	&	\nodata	&	\nodata\\
4417	&	4417a	&	4417cl1\tablenotemark{(2)}, cl2	\tablenotemark{(3)}	&	3	&	Y\tablenotemark{$\dagger$}	&	\nodata	&	\nodata\\
	&	4417b	&	4417cl3	\tablenotemark{(4)}	&	2	&	N	&	05379+3550	&	$3.6\times10^3$	\\
	&	4417c	&	4417cl4	\tablenotemark{(4)}	&	2	&	Y\tablenotemark{$\dagger$}	&	05382+3547	&	$2.0\times10^3$\\
4423	&	\nodata	&	4423cl1	\tablenotemark{(1)}	&	4	&	N	&	\nodata	&	\nodata	\\
	&	4423a	&	4423cl2	\tablenotemark{(2)}	&	2	&	Y\tablenotemark{(6)(7)}	&	05375+3540	&	$1.3\times10^4$\\
	&	\nodata	&	4423cl3	\tablenotemark{(1)}	&	4	&	N	&	\nodata	&	\nodata	\\
	&	4423b	&	4423cl4	\tablenotemark{(1)}	&	3	&	N	&	05375+3536	&	$8.9\times10^3$\\
4565	&	4565a	&	4565cl	\tablenotemark{(2)}	&	3	&	N	&	06155+2319A	&	$5.0\times10^3$\\
	&	4565b	&	4565cl	\tablenotemark{(2)}	&	3	&	N	&	\nodata	&	\nodata\\
	&	4565c	&	N		&	1	&	N	&	\nodata	&	\nodata\\
4678	&	4678		&	N		&	1	&	N	&	\nodata	&	\nodata\\
4753	&	4753a	&	4753cl1\tablenotemark{(2)}, cl2	\tablenotemark{$\dagger$}	&	2	&	Y\tablenotemark{(6)(8)}	&	06382+0939	&	$4.7\times10^2$\\
	&	4753b	&	4753cl3	\tablenotemark{(2)}	&	2	&	Y\tablenotemark{(6)(8)}	&	06384+0932	&	$2.5\times10^3$\\
4974	&	4974		&	4974cl	\tablenotemark{(2)}	&	2	&	Y\tablenotemark{(6)}	&	06053-0622	&	$4.2\times10^4$\\
\enddata

\tablenotemark{a} `Y' indicates detection and `N' indicates no detection. 

\tablenotemark{b} Selected $IRAS$ point source using the selection criteria proposed by \cite{Dobashi1994}.

\tablenotemark{c} The bolometric luminosity calculated following the method of \cite{Myers1987}. 

\tablecomments{\tablenotemark{$\dagger$} The IR clusters or outflows identified by this work.}

\tablerefs{
\tablenotemark{(1)} \cite{Bica_1}
\tablenotemark{(2)} \cite{Bica_2}
\tablenotemark{(3)} \cite{Camargo2011}
\tablenotemark{(4)} \cite{Kirsanova2008}
\tablenotemark{(5)} \cite{Nakano2017}
\tablenotemark{(6)} \cite{Wu}
\tablenotemark{(7)} \cite{Nakano1986}
\tablenotemark{(8)} \cite{Buckle4753}
}

\end{deluxetable*} 
\clearpage


\begin{deluxetable*}{cccccc} 
\tablecaption{Molecular Constants \label{tab:param}}
\tabletypesize{\scriptsize} 
\tablewidth{0pc} 
\tablehead{ 
\colhead{Molecule}	&	\colhead{Transition}	&	\colhead{$S_{ij}$}	&	\colhead{$B_{0}$} 	&	\colhead{$\mu$}	&	\colhead{$E_u$}	\\
\colhead{}	&	\colhead{}	&	\colhead{}	&	 \colhead{(GHz)}	&	\colhead{(Debye)} 	&	\colhead{(cm$^{-1}$)} 	
}
\startdata
CCS			&	$J_{N}$=4$_{3}-3_{2}$	&	3.972	&	  6.477750	&	2.88000	&	3.75568	\\
HC$_{3}$N	&	$J=5-4$				&	5.000	&	  4.549059	&	3.73172	&	4.55219	\\
C$^{34}$S	&	$J=2-1$				&	2.000 	&	24.103540 	&	1.95700 	&	6.19802 	\\
CS			&	$J=2-1$				&	2.000 	&	24.495560 	&	1.95700 	&	4.90249 	\\
SO			&	$J_{N}$=2$_{3}$--1$_{2}$&	2.934 	&	21.523556 	&	1.53500 	&	6.41219 	\\
C$^{18}$O	&	$J=1-0$				&	1.000 	&	54.891420	&	0.11049 	&	3.66194 	\\
$^{13}$CO	&	$J=1-0$				&	1.000 	&	55.101010	&	0.11046 	&	3.67592 	\\
$^{12}$CO	&	$J=1-0$				&	1.000 	&	57.635960	&	0.11011 	&	3.84503 	\\
\enddata

\tablecomments{All of the constants are taken from Splatalogue.}

\end{deluxetable*} 

\clearpage


\begin{figure*}
\begin{center}
\includegraphics[scale=0.4]{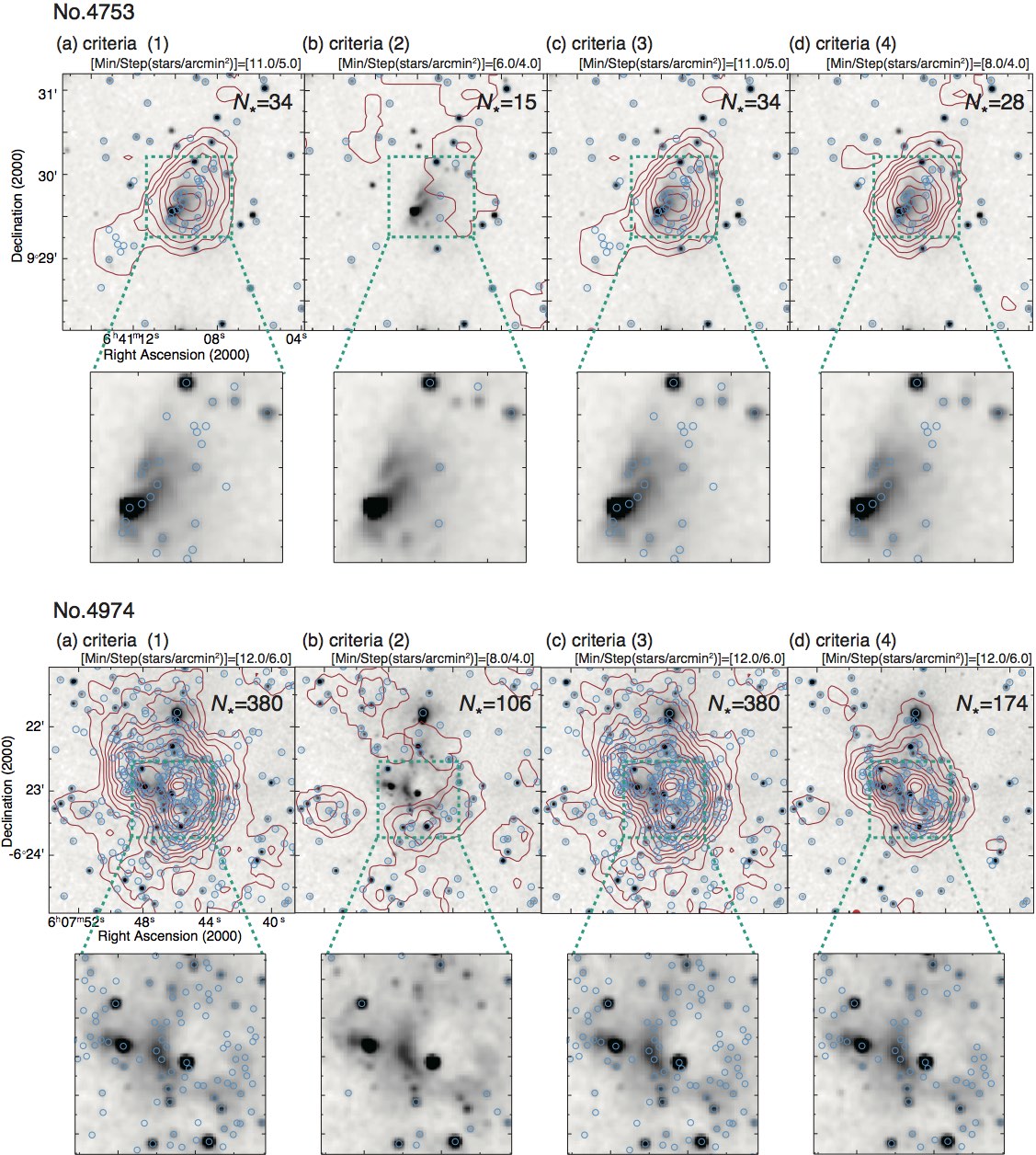}
\caption{
Distributions of stars (blue circles) in the No.4753 and No.4974 regions
selected under the selection criteria (1)--(4) in Appendix \ref{sec:cluster_param}. 
Star densities are shown by the red contours.
Close-up views of the central regions are also shown in the lower panels.
The background is the 2MASS $J$ band image.
 \label{fig:stardensity}
 }
\end{center} 
\end{figure*}


\clearpage

\appendix
\section{Derivation of the Column Density} \label{sec:column}

The brightness temperature corrected for the main beam efficiency $T_{\rm mb}$ of the observed emission line is given by
\begin{equation}
\label{eq:radiative}
T_ {\rm mb}=[J(T_ {\rm ex})-J(T_{\rm bg})][1-{\rm exp}(-\tau)]\ ,
\end{equation}
where  $\tau$ is the optical depth, $J(T)=T_{0} /(e^{T_{0} /T}-1)$, and $T_{\rm bg}=2.7$ K. $T_{0}$ is a constant calculated as
$T_{0}=h\nu_{0} /k$ where $k$, $h$, and $\nu_{0}$ are the Boltzmann constant, the Planck constant, and 
the rest frequency of the observed emission line, respectively.
On the assumption of the Local Thermodynamic Equilibrium (LTE), 
column density for the optically thin lines can be obtained with the following equation \cite[e.g.,][]{Hirahara},
\begin{equation}
\label{eq:column}
N=\frac{3h}{8\pi^{3}}\frac{Q}{\mu^{2}S_{ij}}\frac{e^{Eu/k{T}_{\mathrm{ex}}}}{e^{T_{0}/T_{\mathrm{ex}}}-1}\frac{1}{J(T_{\mathrm{ex}})-J(T\mathrm{_{bg}})}\int \beta^{-1}T{\rm_{mb}}dv.
\end{equation}
Here, 
$Q$ is the partition function approximated as $Q=k\,T_{\rm ex}/h\,B_{0}$ where $B_{0}$ is the rotational constant of the molecule. 
$\mu$ is the dipole moment, $E_{u}$ is the energy of the upper level, and $S_{ij}$ is the intrinsic line strength of the transition for $i$ to $j$ state. 
$\beta$ is the escape probability given by $\beta =(1 - {e^{ - \tau}})/\tau$, and $\beta=1$ for $\tau\ll 1$. 
The spectral line parameters are taken from Splatalogue\footnote{www.splatalogue.net}, 
a database for astronomical spectroscopy.  
We summarize the line parameters used in this study in Table \ref{tab:param}.


\section{Search for Outflows and Estimation of the Outflow Parameters} \label{sec:outflow_param}

In order to search for reliable candidates of outflows,
we checked the \co spectra if they show blue- and/or redshifted high velocity wings around 
the center velocity of C$^{18}$O at each pixel.
We also investigated the $^{12}$CO distributions based on the PV diagrams
which is useful to find outflows and to determine the velocity range of the outflowing gas. 
In total, we found nine outflows which are shown in Figure \ref{fig:outflow}.
We also show some emission lines observed toward the peak position of the high velocity emission.

In Table \ref{tab:outflow}, we provide an estimate of the masses of the outflows $M_{\rm lobe}$ and their properties.
We measured the peak positions of the lobes, and also defined the surface area of the lobes $S_{\rm lobe}$
at the $3\,\sigma$ contour level of the maps for each outflow.
$R_{\rm lobe}$ is the radius of the lobes calculated as $R_{\rm lobe}=\sqrt{S_{\rm lobe}/\pi}$, and
$V_{\rm sys}$  is the systemic velocity defined as the peak velocity of the C$^{18}$O spectra. 
The maximum velocity shift of the $^{12}$CO emission from the systemic velocity,
which we call characteristic velocity $V_{\rm char}$, 
is defined as $V_{\rm char}=\vert V_{\rm sys}-V_{\rm max}\vert$ where $V_{\rm max}$ is the maximum velocity
of the high velocity wings detected at the $3\,\sigma$ noise level.
We tried to measure the optical depth of $^{12}$CO from the ratio of the $^{12}$CO intensity to the $^{13}$CO intensity.
However, there is very little detectable $^{13}$CO emission above the noise level within the lobes.
We therefore measured the upper limit of the optical depth at the intensity peak position of the lobes, $\tau_{\rm max}$, 
and derived properties of the outflowing gas for the upper limits. 

To estimate $\tau_{\rm max}$, we assumed
that the $^{13}$CO is optically thin and that the  [$^{12}$CO]/[$^{13}$CO] abundance ratio is equal to the terrestrial value (89) as follows,
\begin{equation}
\label{eq:tau}
\frac{I(^{12}\mathrm{CO}_{\rm wing})}{I(^{13}\mathrm{CO}_{\rm wing})}=89 \times \frac{1-e^{-\tau_{\mathrm{max}}}}{\tau_{\mathrm{max}}},
\end{equation} 
where  $I(^{12}\mathrm{CO}_{\rm wing})$ and $I(^{13}\mathrm{CO}_{\rm wing})$ are the integrated intensity of the $^{12}$CO and $^{13}$CO emission lines measured at the peak positions of the lobes, respectively.

As the lower limits for the properties of the outflowing gas, 
we assumed that wing components in the \co emission line are optically thin ($\tau\ll 1$).

Using the calculated $T_{\rm ex}$ (see Section \ref{sec:clump}), we derived the $^{12}$CO column densities of the outflows $N(^{12}$CO) 
from Equation (\ref{eq:column}). 
For the estimation of the upper and lower limits, we used $\beta =(1-{e^{-\tau_{\rm max}}})/\tau_{\rm max}$ and $\beta=1$, respectively.
The molecular column density $N$(H$_2$) was then calculated from $N(^{12}$CO), 
assuming a CO fractional abundance of $1\times10^{-4}$ \citep{Frerking}.
Masses of the outflow lobes $M_{\rm lobe}$ were derived from Equation (\ref{eq:mass}) within $S_{\rm lobe}$. 

We list in Table \ref{tab:outflow} the derived $M_{\rm lobe}$ as well as 
the momentum $P_{\rm lobe}=M_{\rm lobe} V_{\rm char}$, 
dynamical timescale $t_{\rm d}=R_{\rm lobe}/V_{\rm char}$, 
and momentum supply rate $\dot{P}_{\rm lobe}=P_{\rm lobe}/t_{\rm d}$.
In this study, we made no attempt to correct for the inclination angle of the outflows.


\section{Criteria to select stars Based on the 2MASS PSC}\label{sec:cluster_param}
In order to search for IR clusters in this paper, we used all of the stars in the 2MASS PSC except for those
identified as minor planets,
because most of the cataloged stars are significantly detected at least in one of the
2MASS bands ($JHK{\rm s}$). However, as pointed out by \citet{Cambrsy2002},
it is known that there are a certain fraction
of false stars in the 2MASS PSC around bright stars and nebulae
associated with small H{\sc ii} regions.
In order to check how much our sample of stars can be contaminated by the false stars,
we made a test by applying some tighter criteria and by checking the appearance of the selected stars
on the 2MASS images.

The test was made toward the dust condensations No. 4753 (NGC 2264) and No. 4974 (Mon R2) that are known to be accompanied by IR clusters \cite[e.g.,][]{Peretto,Dierickx}. 
We produced star density maps covering a $15\arcmin \times15\arcmin$ area around No. 4753 
and a $40\arcmin \times40\arcmin$ area around No. 4974 
using stars selected under the following four different criteria: 

(1) Stars except for those with a minor planet flag. \\
\,\,\, (2) Stars satisfying the above (1), and having a photometry quality flag of A, B, or C and a read flag of 1, 2, or 3 in all of the three $JHK$s bands. \\
\,\,\, (3) Stars satisfying the above (1), and having a photometry quality flag of A, B, or C and a read flag of 1, 2, or 3 at least in one of the $JHK$s bands. \\
\,\,\, (4) Stars satisfying the above (1), and having a catalogued magnitude of $J <16.0$ mag, $H <15.5$ mag, 
and $K$s$<15.0$ mag. 

The criteria (1) are what we used in this study, 
the criteria (3) are to ensure the detection at least in one of the 2MASS bands, 
the criteria (4) are tighter than these, and the criteria (2) are the tightest.

The results are summarized in Figure \ref{fig:stardensity} 
where the individual selected stars (blue circles) and star densities (red contours) of the two regions are shown. 
The 2MASS $J$ band image is overlaid in all of the panels. 
We estimated $N_\star$ the number of stars in the two clusters by removing the background based on the $3 \sigma$ clipping procedure in the same way as described in Section \ref{sec:cluster}. 
The values of $N_\star$ are given in the upper-right corner of each panel. 

As seen in the figures, in the case of the criteria (2),
there are many stars apparently escaped selection because the criteria are too tight, 
and we cannot even realize the existence of the cluster in the case of No. 4753.
In the case of the criteria (3), almost the same stars are selected as the criteria (1). 
In the case of the criteria (4),
the spatial distributions of the clusters (traced by the contours for the star densities)
are similar to those derived using the criteria (1),
though there are still some apparent real stars which escaped selection 
and $N_\star$ decreases by $\sim20\%$ and $\sim60\%$ for Nos. 4753 and 4974, respectively.

In the lower panels of Figure \ref{fig:stardensity}, we also show a close-up view of the central regions.
There are some stars among those selected with the criteria (1) and (3) which we cannot determine if they are real stars or false detections due to the influence of nearby bright source and lumpy/filamentary structures of the nebula. 
However, the criteria (1) allow us to select all of the apparent stars in the $J$ band images, 
while the criteria (2) and (4) miss many obviously real stars. 

In addition to the above (1)--(4), we also tried the criteria that \citet[][see their Section 3]{Cambrsy2002}
employed to identify IR clusters in the North America and Pelican nebulae, and found that
the resulting stars are similar to those selected by the above criteria (2).

We believe that there is no ideal criteria which can exclude false stars perfectly from the 2MASS PSC
in all of the IR clusters, and thus we have to choose realistic criteria to better measure the cluster
parameters we want to know. 
Because we first need to know the existence of clusters and secondly their distributions and extents even roughly, 
we decided to use the criteria (1) in this paper. 

\end{document}